THE SEARCH FOR HABITABLE WORLDS: 1. THE VIABILITY OF A STARSHADE MISSION

Short title: Searching for Earths with a Starshade

Margaret C. Turnbull[1], Tiffany Glassman[2], Aki Roberge[3], Webster Cash[4], Charley Noecker[5], Amy Lo[2], Brian Mason[6], Phil Oakley[7], & John Bally[4]

[1] Global Science Institute, P.O. Box 252, Antigo, WI 54409; turnbull.maggie@gmail.com
[2] Northrop Grumman Corp, One Space Park Drive, E1-4068, Redondo Beach, CA 90278
[3] Exoplanets and Stellar Astrophysics Laboratory, NASA Goddard Space Flight Center, Greenbelt, MD 20771
[4] Center for Astrophysics and Space Astronomy, Campus Box 389, University of Colorado, Boulder, CO 80309
[5] Civil Space Systems, Ball Aerospace & Technologies Corp, 1600 Commerce St, M/S RA4, Boulder, CO 80301
[6] U.S. Naval Observatory, 3450 Massachusetts Ave NW, Washington, DC 20392
[7] Massachusetts Institute of Technology, 77 Massachusetts Ave, 37-582BB, Cambridge, MA 02139

ABSTRACT

As part of NASA's mission to explore habitable planets orbiting nearby stars, this paper explores the detection and characterization capabilities of a 4-m space telescope plus 50-m starshade located at the Earth-Sun L2 point, a.k.a. the New Worlds Observer (NWO). Our calculations include the true spectral types and distribution of stars on the sky, an iterative target selection protocol designed to maximize efficiency based on prior detections, and realistic mission constraints. We carry out both analytical calculations and simulated observing runs for a wide range in exozodiacal background levels ($\varepsilon = 1 - 100$ times the local zodi brightness) and overall prevalence of Earth-like terrestrial planets ($\eta_\oplus = 0.1 - 1$). We find that even without any return visits, the NWO baseline architecture (IWA = 65 mas, limiting FPB = $4\times10^{-11}$) can achieve a 95% probability of detecting and spectrally characterizing at least one habitable Earth-like planet, and an expectation value of ~3 planets found, within the mission lifetime and $\Delta$V budgets, even in the worst-case scenario ($\eta_\oplus = 0.1$ and $\varepsilon = 100$ zodis for every target). This achievement requires about one year of integration time spread over the 5 year mission, leaving the remainder of the telescope time for UV-NIR General Astrophysics. Cost and technical feasibility considerations point to a "sweet spot" in starshade design near a 50-m starshade effective diameter, with 12 or 16 petals, at a distance of 70,000-100,000 km from the telescope.

KEYWORDS: solar neighborhood – astrobiology – telescopes – planetary systems – zodiacal dust – stars: solar type

# 1. INTRODUCTION

## 1.1 *The Starshade*

As a topic in which scientists and the general public share an intense interest, the next major performance goal in the study of exoplanets is to directly image and characterize Earth-like planets orbiting nearby stars (National Academy of Sciences 2010[1]; Lunine et al. 2008). The capability of planned missions to spectroscopically probe the compositions of exoplanetary atmospheres and look for surface signatures is key to discovering whether habitable planets and life may be common in the nearby universe (Woolf et al. 2002; Turnbull et al. 2006; Tinetti et al. 2006). The great challenge underlying this goal is the need to suppress the light of the central star by a factor of $\sim 10^{10}$ or more, while still allowing light from planets to be detected at angular separations of ~0.1 arcseconds or less from the star.

Until recently, the best strategies for doing this seemed to be infrared interferometry (Lawson & Dooley 2005; Lawson et al. 2007), or, in the optical, internal coronagraphs (Levine et al. 2006). However, after a decade of significant investment in internal coronagraphs and free-flying or fixed-beam interferometer concepts for a planet-finding and characterizing mission (e.g., NASA's TPF-C, TPF-I and ESA's Darwin), such designs continue to struggle with great technological complexity and limited usefulness in other kinds of astrophysical observations. Two criticisms that internal coronagraphs and interferometers have faced are: (1) they have large optical systems with requirements on wavefront sensing accuracy and stability that are difficult to achieve (and in the case of interferometers, difficult to test) before launch, and (2) various design factors limit their ability to accommodate the research objectives of the wider astronomical community. In the case of internal coronagraphs, the use of stops and masks causes heavy loss of throughput and drives trades on other throughput factors, e.g., mirror coatings. Consequently, observation at UV wavelengths (a regime which is of great astrophysical interest) may be especially challenging with an internal coronagraph.

In this paper, we explore another idea for directly detecting and studying exoplanets which may alleviate some of the technological difficulties as well as expand the opportunities for partnership with the general astrophysics and cosmology communities: the external occulter, or "starshade". In this architecture (illustrated in **Figure 1)**, a starshade is maneuvered into the line of sight between the telescope and target star, projecting a deep stellar shadow onto the telescope while allowing the light from sources near the star to pass. This relatively simple design (discussed in greater detail in §3) separates the telescope and instruments from the mechanism that blocks the starlight, thereby preventing starlight from ever entering the optical train where internal scattering can greatly complicate observations.

Because starlight is blocked before it enters the telescope, a starshade allows the use of a generic diffraction-limited telescope, without the issues of precision wavefront control that are the central issue for internal coronagraphs (Shaklan et al. 2006). The concept of an "outer" working angle is thereby eliminated, and whole planetary systems can be studied at once, out to an arcminute radius or more. Furthermore, the optical train can be reduced to just 3-4 mirrors before the instrument, vs. the 20 or more typical of internal coronagraphs. This allows for high throughput (near 50% at 500 nm), making it possible to observe at UV wavelengths (down to at least 120 nm for Al plus $MgF_2$ coatings on the primary and secondary mirrors).

---

[1] http://www.nap.edu/catalog.php?record_id=12951

Finally, because the stellar systems of greatest interest are distributed over the whole sky, a significant fraction of a starshade mission must be spent waiting for the starshade to move large distances to acquire each target. Target scheduling simulations show that, on average, starshade target acquisition will take ~11 days (Glassman et al. 2011). These extended windows of primary science "down time" allow opportunities for other astrophysical observations. Though designed to study habitable planets, a starshade mission could expand the discovery space throughout astrophysics, from the nearby stars to the very distant universe.

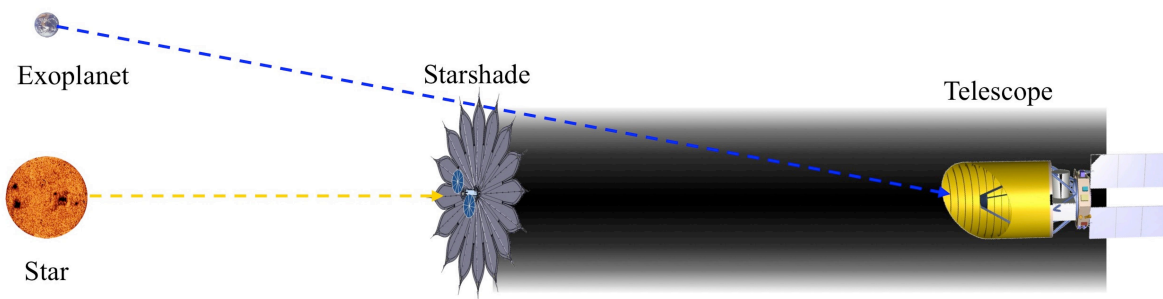

**Figure 1**. A starshade throws a deep stellar shadow over the telescope, but allows light from planetary companions to pass.

## 1.2 *The New Worlds Observer Baseline Design*

In 2008, NASA commissioned an Astrophysics Strategic Mission Concept Study[2] for the New Worlds Observer (NWO), a telescope plus starshade observatory designed to provide (1) efficient detection and characterization of Earth-like and larger exoplanets in the habitable zones of nearby stars and (2) a revolutionary program of general astrophysics. In this paper, we build on the results of that study and ask: What are the starshade and telescope parameters required for NWO to achieve its scientific objectives, given the inevitable and (sometimes unknown) constraints? To answer this question we must translate science objectives into performance requirements, and these requirements must be expressed in terms of feasible telescope and starshade engineering parameters.

**Table 1** lists the NWO science objectives and major constraints used in this paper to define the baseline mission and engineering parameters. The primary science objective of NWO is to achieve a 95% probability of detecting and spectrally characterizing at least one Earth-like planet (where "Earth-like" is defined in terms of albedo and size), assuming that the prevalence of such planets in the solar neighborhood is 10% or greater ($\eta_\oplus \geq 0.1$) and that the typical exozodiacal background is no more than 100 times that of the Solar System ($\varepsilon \leq 100$). Thus we are seeking to match or exceed the requirements described in the TPF-C STDT Report (Levine et al. 2006).

Table 1
NWO Objectives and Constraints

| Science Objectives |
| --- |
| 1. Achieve at least a 95% probability of detecting at least one habitable Earth-like planet, assuming $\eta_\oplus \geq 0.1$ and $\varepsilon \leq 100$, and realistic background sources |
| 2. Characterize spectral features such as water, oxygen and ozone |
| 3. Reserve at least 50% of mission lifetime for General Astrophysics (GA) |

| Constraints |
| --- |
| 1. Complete objectives within a 5 year mission lifetime |
| 2. Maintain a technically feasible mission architecture (with appropriate engineering constraints such as solar avoidance, $\Delta V$ limits, etc.) |

As discussed by Levine et al. (2006), the requirement of achieving a 95% probability of at least one habitable Earthlike planet detection can be expressed as a minimum number of habitable zones that must be searched. For example, in a program where 100% of each target's habitable zone is within the detection space of the observatory, the expectation value for the number of potentially habitable planets detected is approximately $\left\langle N_{php} \right\rangle = \eta_\oplus \times N_{targets}$, assuming a maximum of one habitable planet per system and that each system is observed one time (as assumed here). Thus for a program of 30 stars, the expectation value for the number of habitable planets found would be 3. If such a search of 30 target stars revealed no planets, the probability that $\eta_\oplus \geq 0.1$ would be $0.9^{30}$, or about 4%. Conversely, the probability of detecting at least one habitable Earth-like planet for such a program, assuming $\eta_\oplus \geq 0.1$, is $1\text{-}0.9^{30}$, or greater than 95%. A mission designed to accomplish this goal – taking into account the distances, spectral types, and particular spatial distribution of stars in our sky – will either revolutionize the field of exoplanetary studies, or in the case of no detections, it will provide a strong upper bound on $\eta_\oplus$ and on the prevalence of life as we know it in the Solar Neighborhood.

Because the set of nearby stars includes a variety of spectral types at different distances, the habitable zone coverage for each star will be somewhere between 0 and 100%. Therefore, as investigated in this paper, achieving the NWO science objectives will require more than 30 targets. Furthermore, as stated in Table 1, the mission must be designed to handle realistic background sources. Depending on the density of faint sources near the most favorable targets, this may mean that multicolor or spectroscopic observations are required in every case. We explore that scenario briefly in the following sections, but we note that the topic of background sources is in need of more detailed investigation in the future.

A feasible fuel carrying capacity for both telescope and starshade points to a 5-year mission lifetime (assuming no servicing missions; Cash et al. 2009). Since the primary objectives must be suspended while the starshade is in transit between primary science (exoplanet) targets, 50-70% of the mission's lifetime is available for General Astrophysics (GA) programs. To accommodate a wide array of astronomy programs, a minimum of fifty percent GA time over a 5 year mission is the NWO baseline requirement stated in Table 1.

## 1.2.1 *Telescope Size*

The scientific objective to achieve a 95% probability of detecting at least one Earth-like planet within a nominal 5-year mission lifetime drives the telescope diameter, via both (1) the integration time required to detect and characterize exoplanets and (2) the spatial resolution required to avoid confusion with exozodiacal light.

**Figure 2** illustrates, from left to right, the face-on Solar System at 10 pc distance, imaged with 2-m, 4-m and 8-m telescopes in combination with a starshade providing a 65 milliarcsecond inner working angle (IWA). The right-most image also shows a wider field of view for the 4-m case, showing planets and exozodiacal dust structure out to the orbit of Neptune. The input Solar System model used for these simulated images includes a high-fidelity model for the Solar System's debris dust (Roberge et al. 2011, in prep). The inner portion of the dust model (clearly visible interior to Jupiter's orbit) is based on COBE DIRBE observations (Kelsall et al. 1998), while the outer portion (visible in the wider field of view of the right-most panel) is a new theoretical model combining N-body integration with a collisional algorithm (Kuchner & Stark 2010). The complete Solar System model including exozodiacal dust and planets was subsequently passed through an NWO optical train simulator for the different telescope diameters (T. Glassman, private communication). Note that these images are intended to illustrate the capability of different telescope sizes in terms of relative flux of the various components and imaging resolution, but these images do *not* include noise. The apparent graininess in these images is due to the finite number of particles modeled in the N-body code.

In Figure 2, Jupiter and Saturn are clearly visible with all three telescope diameters (seen near 1 o'clock and 10:30, respectively). The inner planets Venus and Earth (7:30 and 3:00, respectively) are readily detectable just outside the IWA with the 4- and 8-m apertures, but with the lower resolution of the 2-m telescope the light from these planets nearly lost in the bright (and potentially non-uniform) exozodiacal background. Mars is faintly visible in the 8-m case (4:00), even less convincing in the 4-m case, and apparently not detectable at all with the 2-m.

In this paper, we assume a baseline telescope size of 4 meters. As explored in the next Section, a 4-m telescope appears to provide sufficient sensitivity to detect and spectroscopically characterize an Earth in a Solar System twin at 10 pc in ~1 day.

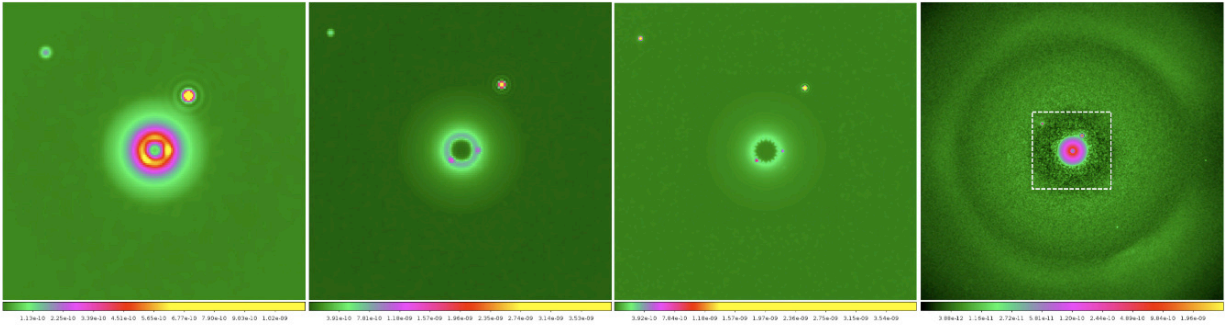

**Figure 2**. Simulated images of the Solar System seen face-on, at a distance of 10 pc, with 2-m, 4-m, and 8-m telescopes (left to right) plus a 50-m effective diameter starshade flying at 80,000 km separation, providing a 65 milliarcsecond IWA. Venus (at 7:30), Earth (3 o'clock), Jupiter (1 o'clock) and Saturn (10:30) are detectable in all three images (though signal from Venus and Earth may be indistinguishable from that of dust structures in the 2-m case), while Mars (4:00) is visible only in the 8-m image. A wider field of view for the 4-m case is shown in the right-most panel, where Uranus (5:00), Neptune (3:00), and exozodiacal dust structures are visible.

### 1.2.2 *Field of view*

As seen in the right-most panel of Figure 2, outer planets can give rise to dust structures that should be detectable with a 4-m telescope, and this is one area of study where the starshade architecture may offer unparalleled capability. In contrast to coronagraphs and interferometers, a starshade has no intrinsic outer working angle, and this enables efficient study of the entire planetary system architecture within a single image. A 25 arcsec$^2$ field of view is adequate to image the alpha Centauri system out to 25 AU, and for most other targets this FOV would be sufficient to image extrasolar Kuiper belts and beyond in a single exposure.

### 1.2.3 *Wavelength coverage and spectral resolution*

In order to accomplish the exoplanet and GA science goals, NWO is envisioned to have near-UV, optical, and near-IR wavelength coverage, similar to the Hubble Space Telescope. The planet detection bandpass, color filters, and spectral resolution chosen for NWO should (1) maximize detection efficiency and (2) help to quickly distinguish between true planets and unresolved background sources. For Sun-like stars, planet detection is most efficient in the 0.3-0.8 μm range, where photon flux is greatest and silicon-based CCD detectors are most sensitive. Furthermore, this wavelength range includes signatures of oxygen, water, and ozone, all of which are considered indicators of habitability or even extant life (Des Marais et al. 2002). Planet detection near the inner working angle is more difficult at longer wavelengths due to leakage of starlight around the starshade; therefore the starshade engineering specifications must be chosen to achieve the necessary starlight suppression for the longest wavelength of our detection bandpass, 0.8 μm.

Due to the high probability of finding faint unresolved background sources within the NWO field of view, it is important that the initial detection observation be carried out simultaneously in more than one bandpass. True planets detected in reflected starlight will appear to be the same color as the target star but 23-27 magnitudes fainter, which rules out nearly all background sources as convincing planet candidates.

Near the Galactic plane, a highly extincted star could have the V-band brightness expected for a planet, but due to interstellar reddening these objects generally appear far redder than the roughly Sun-like NWO target stars. The exception to this is white dwarfs, which occupy the same (V, B-V) space as planets, but with thin disk space densities of $\sim 10^{-3}$-$10^{-2}$ pc$^{-3}$ (Festin 1998, Liebert et al. 1988, Ruiz & Takamiya 1995, Oswalt et al. 1996, Pirzkal et al. 2005) such objects should be relatively uncommon near the target star's HZ ($N_{WD} \sim 10^{-2}$ arcsec$^{-2}$) and readily distinguished from planets with multi-band photometry. **Figure 3** shows the observed B-V color magnitude space occupied by habitable planets and Jupiter analogs, as compared to extincted white dwarfs, giants, and main sequence stars at distances of 100 pc to 25 kpc, assuming extinction rates between $A_V = 0.1$-10 mag/kpc, and color excesses (B-V) $-$ (B-V)$_0$ = $E_{B-V}$ = $A_V$/3.2.

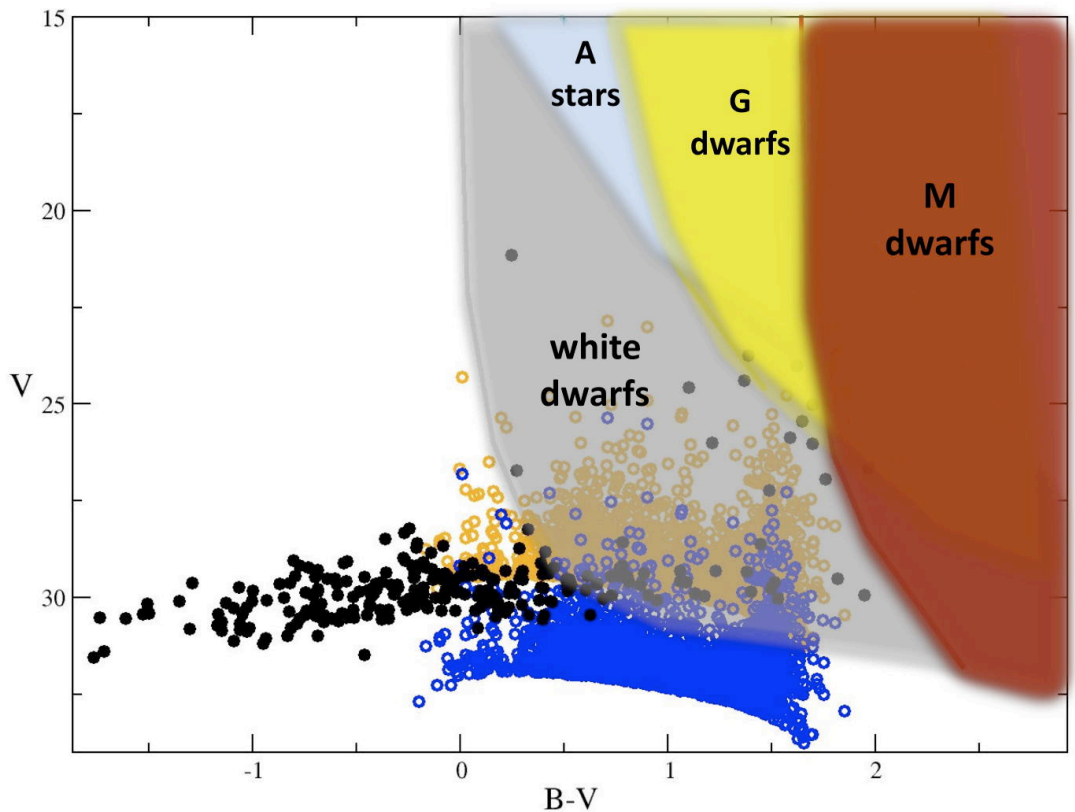

**Figure 3.** The approximate B-V color magnitude space occupied by exoplanets seen in reflected starlight (blue circles=Earths, orange circles=Jupiters) compared to stars (shaded regions) and unresolved background galaxies (black dots). For a background star to masquerade as an Earth-like planet, a high extinction with implausibly little reddening would be required. White dwarfs and faint unresolved galaxies in the B-V range of interest should be relatively rare ($\sim 10^{-3}$ arcsec$^{-2}$), but will require multi-band photometry for disambiguation should they appear near the HZ.

Away from the Galactic plane, NWO will essentially see a Hubble Ultra Deep Field in every image, and the primary concern is a veritable ocean of (1) brighter (V<28) extended galaxies and (2) ultra-faint unresolved galaxies (mostly V>28). **Figure 4** shows a 25 arcsec$^2$ NWO field superimposed on the Hubble Ultra Deep Field (HUDF, Beckwith et al. 2006). We expect to find ~100 extragalactic sources in every NWO field of view, and a resolved galaxy near the target star's HZ in every ~2 fields. These non-uniform sources could make planet detection difficult wherever they dominate the signal. We expect just as many unresolved sources with V > 28 within the central square arcsecond around each target star, but number densities could be higher beyond the HUDF detection limit of V~29 (Windhorst et al. 2008). Fortunately, the color distribution of these unresolved sources tends toward the extreme blue at fainter magnitudes (Coe et al. 2006). Figure 3 shows the HUDF faint unresolved sources as black dots (sources with V > 28, stellarity > 0.9; Coe et al. 2006), and there is likely to be some overlap with Earth and Jupiter analogs in B-V color-magnitude space. Again, multi-band photometry and/or spectroscopy would readily disentangle these sources, as ultra-faint galaxies are observed to have markedly different spectral energy distributions from stars in the optical (Coe et al. 2006; Pirzkal et al. 2005). In the rare case where an unresolved background galaxy falls within the expected multi-color range for planets, spectroscopic follow-up and proper motion discrepancies (~2-20 milliarcseconds/day) will be able to disambiguate these sources within a few days.

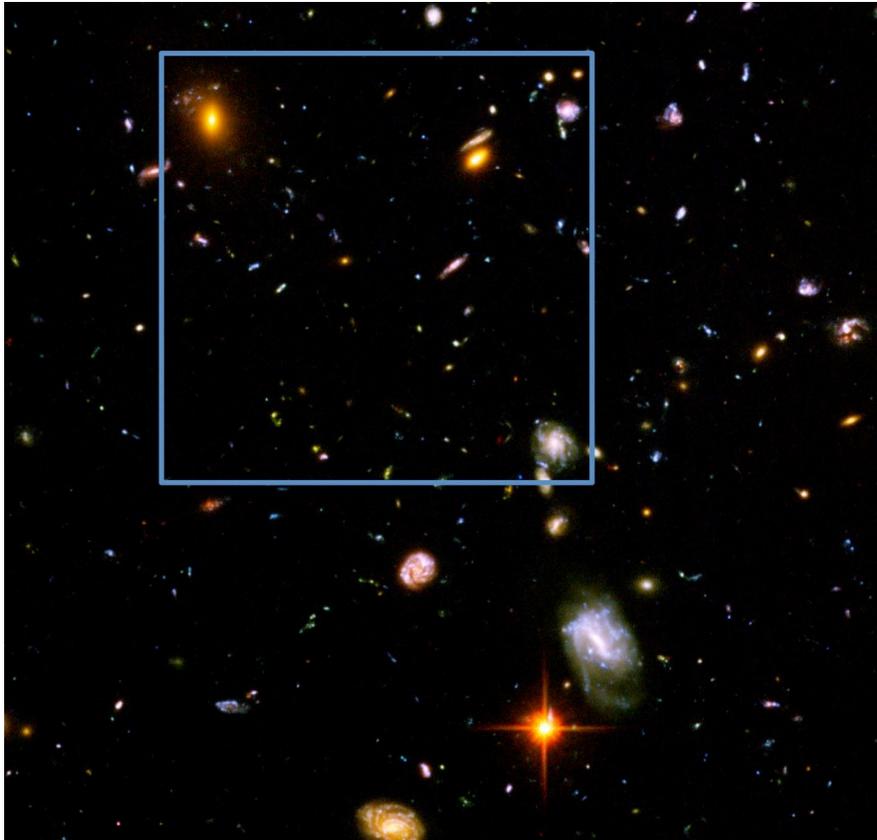

**Figure 4.** A 25 arcsec$^2$ field of view for NWO (blue outline), superimposed on a portion of the Hubble Ultra Deep Field image (Beckwith et al. 2006). In a one day exposure, the NWO baseline mission will detect exoplanets and ~$10^2$ background objects of similar brightness or slightly fainter than the sources in the Ultra Deep Field. Most such objects can be distinguished from planets via multi-band photometry.

Finally, unresolved exozodiacal dust clumps may occasionally be bright enough to masquerade as planets, and this may be the greatest confounding unknown in exoplanet imaging at optical wavelengths. While such clumps would be interesting in their own right, spectroscopic analysis may be the most efficient way to distinguish them from planets.

The precise determination of optimal bandpasses for efficiently distinguishing between planets and background sources remains for future work; for our purposes in this paper it suffices to say that given the inevitable presence of background sources described here, the NWO mission must provide for detection in several bandpasses. If the initial observation is carried out simultaneously in 3 or 4 broadband filters, it will almost always be possible to discriminate between planets and distant background sources in a single visit.

Once planets (or planet candidates) are detected, spectroscopic follow-up can be carried out with longer integration times to rule out dust clumps and characterize any bone fide planets. Spectroscopic analysis involves dividing the spectrum of the planet candidate by the spectrum of the star itself, and this operation results in a reflectance spectrum wherein atmospheric and surface constituents can be identified (cf. Woolf et al. 2002; Turnbull et al. 2006; Montañes Rodriguez et al. 2004; Tinetti et al. 2006). For planets farther out from the IWA, spectroscopy will be possible at longer wavelengths (ideally, out to 1.6 microns for detection of deeper water bands, methane, and $CO_2$) by flying the starshade in closer to the telescope, and suppressing longer wavelength diffraction at the expense of the IWA (Cady et al. 2009). Wavelength coverage in the ultraviolet is also necessary for many General Astrophysics applications, and efficiency at these wavelengths is largely dependent on mirror coatings. One possible combination is $MgF_2$ over aluminum, which provides good reflectivity from near-IR to ~120 nm (Quijada et al. 2010). Other coatings (e.g., LiF over aluminum; Raymond et al. 2000) can provide better efficiency at shorter wavelengths but at greater cost due to their fragile nature.

### 1.2.4 *IWA and FPB*

The achievable limits on inner working angle and fractional planet brightness (FPB) are the heart and soul of any mission to directly image and spectrally characterize habitable planets. Together, the IWA and limiting FPB determine which stars' habitable zones can be searched for planets, and both parameters depend critically on the starshade size, shape and distance from the telescope. Improving the mission performance in either of these areas requires a larger, more distant starshade, which requires longer slew time between targets and a greater fuel payload (Arenberg et al. 2008, Vanderbei et al. 2007). Therefore it is in the interest of the mission to minimize the size and distance of the starshade and to work with the largest IWA and limiting FPB that can accomplish the science goals.

### 1.2.5 *Enabling General Astrophysics*

Finally, an essential requirement is that NWO represent a truly revolutionary tool for a wide range of general astrophysical (GA) studies, the possibilities for which we discuss in Section 3. Among other considerations, this means providing (1) sufficient wavelength coverage and (2) sufficient observing time to carry out such a program. As discussed above, the vastly simplified optical train of a starshade architecture makes observations from the UV through near-IR feasible, and NWO will have angular resolution nearly a factor of two better than Hubble or JWST. To ensure that substantial time is reserved for GA observations, in Section 2 we assess the primary science performance within the context of a GA program that takes 50-70% of the mission lifetime.

In Section 2, we determine the necessary IWA and FPB for a viable mission, following these steps: First, we calculate the angular sizes of habitable zones around stars within 30 parsecs and the brightness of Earth-like planets orbiting within those habitable zones (§§2.1-2.3). From this we estimate the baseline performance requirements on inner working angle and limiting fractional planet brightness that will enable NWO to achieve the science goals (§2.4). Using these performance requirements we calculate a "completeness" value for each star (§2.5), and we use the fractional planet brightness at the Earth-equivalent insolation distance (EEID; 1 AU for the Sun) to calculate the exposure times required to detect and spectrally characterize an Earth-like planet assuming a wide range of exozodiacal dust levels (§2.6).

Given these completeness values and exposure times for the baseline NWO mission, in §2.7 we analyze mission performance in terms of number of Earths detected and characterized using (1) a simple analytical code to prioritize stars for observation and to examine the effect of exozodiacal light on mission performance, and (2) a more sophisticated Monte Carlo scheduling simulation that accounts for additional engineering constraints and the real distribution of stars on the sky. These scheduling simulations reveal the most favorable targets for observation and the minimum mission lifetime required to observe them, given a range of planet frequencies (10% to 100%), exozodiacal background ($\varepsilon = 1$-100), and realistic slew times between targets. In Section 3 we translate the baseline performance requirements into a physical design for NWO, and in Section 4 we summarize our results and outline the preparatory astrophysics and engineering tasks that must be achieved in order to build a successful starshade mission.

## 2. IMAGING HABITABLE EXOPLANETS

### 2.1 *The Target Stars*

Assessing NWO's ability to accomplish the primary science objectives requires knowledge of the local landscape of stars in terms of stellar spectral types and distances. Due to the high diversity and low density of stars in the Solar Neighborhood, the canonical "Sun at 10 parsecs" is of limited usefulness in assessing the viability of a mission to directly detect Earths. In this Section we examine the nearby stars in terms of the inner working angle and limiting fractional planet brightness required to achieve the goals in Table 1.

We begin with the sample of Hipparcos stars within a distance of 30 parsecs (parallax > 33.33 mas); the parallax and photometry data used for the following calculations are taken from the Hipparcos Catalogue (ESA 1997). There are 2350 Hipparcos stars within 30 pc (including ~200 stars within 10 pc, ~1000 stars within 20 pc), the largest portion of which are main sequence K stars. At least 950 stars in this list are part of a double or multiple system (either physical or optical companions, according to the Washington Double Star Catalog maintained at the USNO). Hipparcos' limiting magnitude is V~12, but the Catalogue is *complete* only down to V=7.3-9 (depending on star color; ESA 1997). As a result, the 30 pc sample is complete for G-, F-, A- and B-type stars and giants, but not for the fainter K- and M-stars or white dwarfs.

Stellar luminosity was estimated for Hipparcos stars using Hipparcos parallax data, V-band bolometric corrections interpolated from Flower (1996) and Hipparcos Johnson V and B-V values (corrected according to Bessel 2000, Turnbull & Tarter 2003a). Hipparcos $H_p$, and Tycho $B_T$ and $V_T$ magnitudes for each star were converted to Johnson B and V magnitudes via polynomial fits to Bessell's (2000) recalculations of the Hipparcos and Tycho passbands. Residuals between our polynomial fits and the Bessell calculations were less than 0.01 magnitudes. The ground-based (B-V) measurement listed in the Hipparcos Catalogue was used when the uncertainty in that measurement was smaller than that of the $(B_T - V_T)$ measurement, and the ground-based (V-I) measurement was used when neither (B-V) nor Tycho photometry were available. Both $H_p$ and $(B_T - V_T)$ were used to calculate Johnson V magnitude, depending on which of the $H_p$ or $(B_T - V_T)$ uncertainties was smaller.

To estimate stellar luminosities for calculating habitable zone location, we used the bolometric corrections as a function of B-V color from Flower's (1996) Table 3. The bolometric corrections were fit to a polynomial in terms of (B-V):

$$
\begin{aligned}
\text{B.C.} = \quad & -2.6703 \, (B\text{-}V)^6 \\
& + 11.92 \, (B\text{-}V)^5 \\
& - 21.088 \, (B\text{-}V)^4 \\
& + 18.552 \, (B\text{-}V)^3 \\
& - 9.1981 \, (B\text{-}V)^2 \\
& + 2.1958 \, (B\text{-}V) \\
& - 0.1494
\end{aligned}
\tag{1}
$$

With this fit to Flower's (1996) data, stars with 0.3 < B-V < 1.4 have residuals |$BC_{fit}$ - $BC_{Flower}$| < 0.01 magnitudes. For stars with B-V > 1.8, we used a constant BC of -5.5 magnitudes (the maximum in Flower's table). As shall be seen in the following Sections, it is due to this large bolometric correction for the reddest stars that proxima Cen appears to be a viable target for NWO, a conclusion we regard with skepticism. The sun's bolometric correction (for B-V = 0.65) determined according to Equation 1 is -0.09, in reasonably close agreement with Cox (2000; $BC_{sun}$ = -0.08). Taking the sun's absolute bolometric magnitude to be $M_{BOL}$ = 4.74 (Cox 2000), luminosities were then calculated as:

$$
L_* / L_{sun} = 10^{-0.4(M_V + BC - 4.74)}
\tag{2}
$$

The broad spectral types ("B", "A", "F", "G", "K", and "M") referred to and used to color-code the figures in this paper are taken from the spectral survey carried out by Houk and co-authors (1975, 1978, 1982, 1988 and 1999). In a few cases where Houk's types are not available or are most likely incorrect given other data (i.e., the calculated bolometric luminosity, Tycho B-V photometry, ground-based V-I photometry, and/or recent effective temperature data via Valenti & Fischer; 2005), the B-V color was used to estimate spectral type instead. Following Cox (2000), spectral types were assigned as follows:

| | | |
|---|---|---|
| B stars: | B-V < 0 | $T_{eff}$ > 10,000 K |
| A stars: | $0 \leq$ B-V > 0.3 | $T_{eff}$ = 10,000 - 7300 K |
| F stars: | $0.3 \leq$ B-V < 0.58 | $T_{eff}$ = 7300 - 5940 K |
| G stars: | $0.58 \leq$ B-V < 0.8 | $T_{eff}$ = 5150 - 5940 K |
| K stars: | $0.8 \leq$ B-V < 1.4 | $T_{eff}$ = 3840 - 5150 K |
| M stars: | B-V $\geq$ 1.4 | $T_{eff}$ < 3840 K |

Luminosity classes ("giant", "main sequence", and "sub-MS" below the main sequence) have been assigned based on location on the HR diagram as shown in **Figure 5**. Luminosity classes thus defined were generally in good agreement with the Houk types where available. We define the "sub-MS" limit below the main sequence to be:

For B-V < -0.1: $M_V > 28$ (B–V) + 5.8         (3)

-0.1 $\leq$ B-V < 1.28: $M_V > 4.8$ (B–V) + 3.5       (4)

1.28 $\leq$ B-V: $M_V > 17$ (B–V) – 12.2       (5)

We define the "giants" limit above the main sequence to be:

$$M_V \leq -10((B-V)-1.4)^2 + 6.5 \tag{6}$$

Again, these labels are intended merely to visually portray in the following Figures which kinds of stars are available for study given NWO performance specifications. Applying these limits to the 2350 stars in the Hipparcos 30 pc sample, there are 2206 "main sequence" stars, 36 "giant" stars, 62 "sub-MS" stars. Aside from a handful of white dwarfs, the "sub-MS" stars have large uncertainties in parallax and/or photometry (shown in black in Figure 5) and are probably main sequence M stars, all of which are too faint to be feasible NWO targets (explained further in §2.2). Fifty-one of the faintest stars (V ~ 10-13) did not have Tycho or ground-based photometry and are not included in this analysis. **Table 2** lists the total number of stars in our sample for each spectral type, the approximate distance to which the sample is complete (assuming a completeness magnitude of V=8 and typical absolute magnitudes of $M_V$ = -1, 2, 3.5, 5, 7, 12 and 0 for B-, A-, F-, G-, K-, M-type and giant stars respectively), the approximate total number of stars that exist within 30 pc, and the completeness percentage for the 30 pc volume ("%$_{HIP30}$").

Table 2
Stars in the 30-pc Hipparcos Sample

| Spec Type | $N_{HIP30}$ | $d_{comp}$(pc) | N(30pc) | %$_{HIP30}$ |
|---|---|---|---|---|
| B | 4 | 30 | 4 | 100 |
| A | 59 | 30 | 59 | 100 |
| F | 243 | 30 | 243 | 100 |
| G | 436 | 30 | 436 | 100 |
| K | 781 | 16 | 936 | 83 |
| M | 678 | 4 | 4640 | 15 |

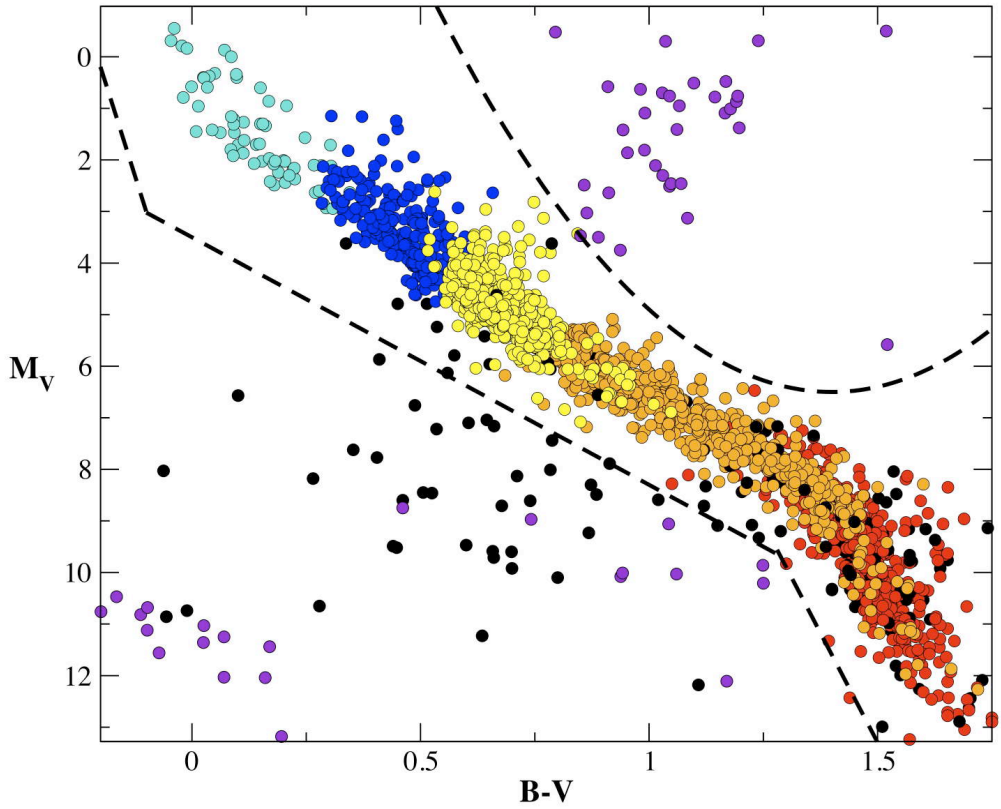

**Figure 5.** The Hipparcos 30 pc sample, color-coded by the spectral type listed in our database: red = M dwarfs, orange = K dwarfs, yellow = G dwarfs, blue = F dwarfs, cyan = A/B stars, purple = giants and "sub-MS" stars. Points in black have B-V and parallax uncertainties greater than 20%. Dashed lines show the divisions between stars we labeled as "giants" (upper right), "main sequence" (center), and "sub-MS" (lower left).

Based on these data, there are a few hundred late-K stars and a few thousand M stars missing from the Hipparcos 30 pc sample (giving a true grand total of ~6,000 stars within 30 pc of the sun). These missing stars are not of interest to the present work; their habitable zones are too small to resolve with a feasible telescope size. On the other hand, some early F- and A-type stars beyond a distance of 30 pc have habitable zones that could be imaged, but because more luminous stars have less favorable planet:star flux ratios (discussed below) there are no viable targets beyond our 30 pc cut-off. It is also important to realize that habitable Earths are all the same absolute brightness, regardless of the star they orbit; therefore more distant targets will require longer integration times. This situation is the same regardless of which starlight suppression technology is used. With the exception of a small handful bright, roughly sunlike stars that were apparently overlooked in the creation of the Hipparcos Catalogue (discussed in Paper 2), we believe the sample of stars studied in this paper includes every viable primary science target for NWO—and indeed for all direct-imaging missions with similar detection limits.

## 2.2 Habitable Zone Angular Sizes and IWA

Determining an appropriate inner working angle for NWO begins with a discussion of the angular sizes of habitable zones surrounding the Sun's neighbors. In this paper we follow convention and define the habitable zone as the range of distances from a star where an Earth-like planet could support liquid water on its surface. While the true extent of this habitable zone depends to some extent on the color (i.e., temperature) of the star and the atmospheric, geological, or even biological processes at work in the planet itself (e.g., Kasting et al. 1993, Catling et al. 2005; Pavlov et al. 2000; Williams et al. 1998), we can simplify the discussion by noting that the empirical evidence (i.e., Venus) indicates that for an Earth-sized planet the inner HZ edge apparently does not extend as close to the Sun as 0.7 AU. The outer edge is less well-constrained but apparently extends beyond 1.5 AU, because Mars appears to have had copious amounts of surface water in the past, and a more massive Mars (with higher surface gravity and carbon cycling) may well have remained habitable to this day. Accounting for the fainter young Sun, the HZ limits for the Sun used here are from 0.75 to 1.8 AU (see Lunine et al. 2008). These limits define the amount of energy a planet can receive to be inside the habitable zone: to calculate the inner HZ (IHZ) and outer HZ (OHZ) limits for other stars, we scale by the square root of stellar bolometric luminosity:

$$\text{Inner HZ Location} = a_{\text{IHZ}} = 0.75\,AU \times \sqrt{L_*/L_{sun}} \qquad (7)$$

$$\text{Outer HZ Location} = a_{\text{OHZ}} = 1.8\,AU \times \sqrt{L_*/L_{sun}} \qquad (8)$$

Translating the linear HZ size (in AU) into an angular size (in mas), we find that the angular HZ size can be expressed in terms of the apparent magnitude alone. The angular Earth-equivalent insolation distance (EEID) is:

$$\vartheta_{HZ}('') = a_{HZ}(AU)/d(pc) = \sqrt{L_*/L_{sun}}\Big/d(pc) \qquad (9)$$

$$= 10^{-(M_{BOL}-4.74)/5}/d(pc). \qquad (10)$$

Taking $M_{BOL} \sim M_V$ and $M_V = V - 5 \log d(\text{pc}) + 5$, a convenient rule of thumb for the habitable zone angular size is:

$$\vartheta_{HZ}('') \approx 10^{-V/5} \qquad (11)$$

Thus for a ~65 mas IWA, a convenient rule of thumb is that *all viable targets must be brighter than V~6*. Given the layout of the Solar Neighborhood, this implies that the latest spectral type searchable with an inner working angle of 65 mas is ~K5V. **Figure 6** shows the angular extent of the habitable zones of Hipparcos stars within 30 pc. Note that this Figure shows HZ locations for the case of pole-on systems. For inclined systems, portions of the HZ will be hidden inside the IWA (Brown 2004, 2005). Considering only HZ angular size, a goal of searching Sun-like stars for Earth-like planets seems achievable with NWO, even though M stars are clearly ruled out (these are the domain of transit searches; see Gould, Pepper & DePoy 2003). Proxima Cen, an M5 dwarf at only ~1pc, may be an exception, but the HZ locations for M stars are highly uncertain due to large bolometric corrections. Using B-V = 1.8, in §2.1 we assigned a BC of -5.5 magnitudes for Proxima Cen, but Frogel et al. (1972) used multi-band photometry to derive a BC of -3.8 magnitudes for this star, which would put the HZ just interior to the range of visibility for NWO.

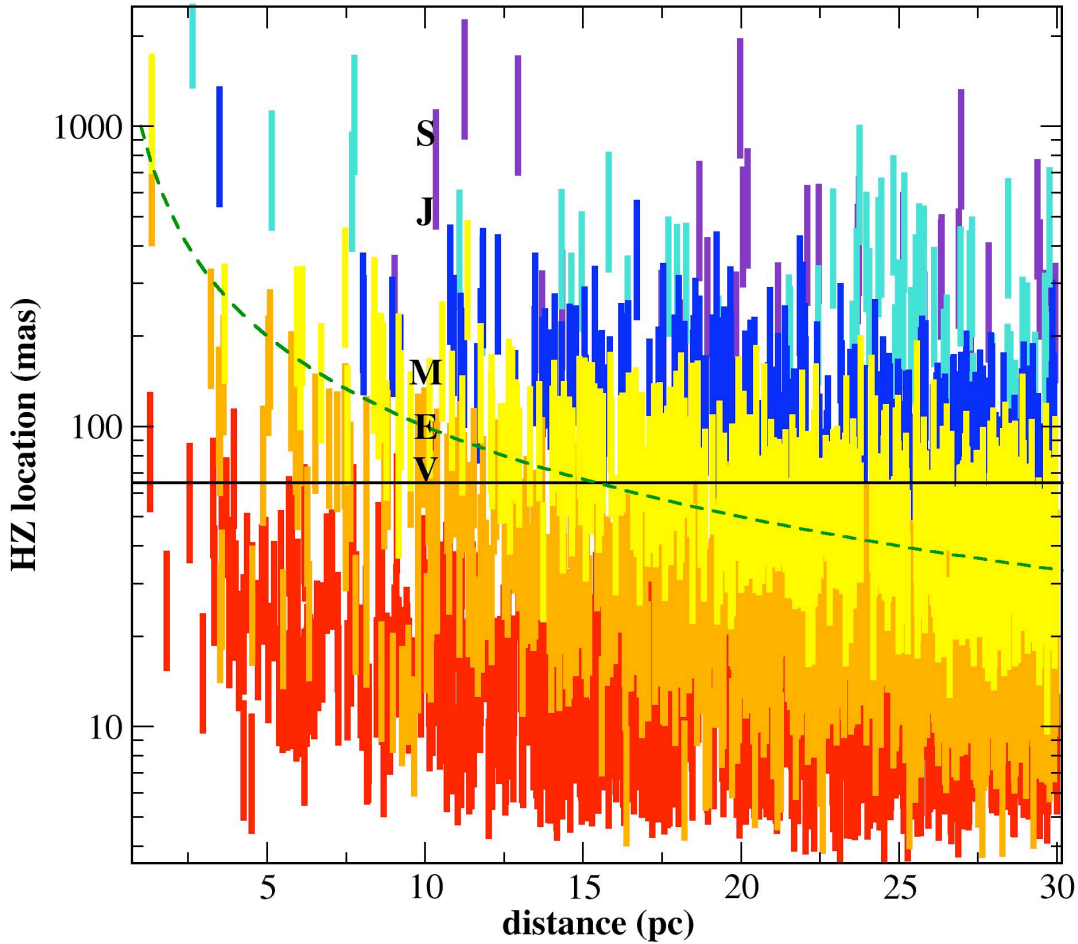

**Figure 6.** Habitable zone locations (0.75-1.8 AU for the sun, seen face-on, scaled for stellar luminosity and distance) for Hipparcos stars within 30 pc, color-coded by spectral type as in Figure 5. An IWA of 65 mas is shown for reference, as is the Earth's angular separation from the sun as a function of distance (green dashed line). The locations of the sun's planets are also shown, scaled for a distance of 10 pc.

### 2.3 *Fractional Planet Brightness*

In addition to being very near to the star, a habitable terrestrial planet is very small in size, and thus reflects only a tiny fraction of the star's light. The amount of starlight reflected by a planet in the habitable zone depends on three things: the planet's radius, its albedo at the wavelength of the detection bandpass, and its location within the HZ. Planet luminosity does *not* depend on the luminosity of the star; by definition, "the habitable zone" is where the planet receives the right amount of energy for an Earth-like planet to have liquid water on its surface (about $10^6$ ergs sec$^{-1}$ cm$^{-2}$), regardless of which star it orbits.

A simple rule is that, for planets of a given size and albedo, planet brightness is roughly constant and the fractional planet:star brightness ratio (or FPB) depends only on the stellar luminosity. As in Brown (2005), the FPB is calculated according to:

$$(F_p / F_{*})_{HZ} = \Phi(\beta) \, r_p^2 \, p / a_{HZ}^2, \tag{12}$$

where $p$ is the geometric albedo and $\Phi(\beta)$ is the phase factor as a function of the star-planet-observer angle, $\beta$. For a Lambertian sphere, $p$ is 2/3 times the Bond albedo ($A_B \sim 0.3$ for the Earth; Goode et al. 2001), and $\Phi(\beta)$ is $1/\pi$ at maximum elongation ($\beta = 90$ degrees), giving for the Earth orbiting the Sun at 1 AU:

$$(F_p/F_*)_{HZ} = 0.2(r_p^2 / \pi a_{HZ}^2) = 1.155 \times 10^{-10}. \tag{13}$$

Using $a_{HZ}(AU) \propto \sqrt{L_* / L_{sun}}$, we find:

$$(F_p/F_*)_{HZ} = \Phi(\beta) r_p^2 p / L_* = 1.155 \times 10^{-10} / L_* \tag{14}$$

at the EEID (which is 1 AU for the Sun). A comparison of (9) and (14) brings home the problem for planet-imaging missions: that angular habitable size goes as stellar luminosity $L_*$, while fractional planet brightness goes as $1/L_*$, and both are quantities that we wish to maximize in choosing target stars.

It must be noted, however, that planet location *within* the habitable zone also affects the planet's brightness, and planets at the outer edge of the HZ are fainter than at the inner edge, by a factor of ~5.8. At the inner and outer HZ edges (0.75-1.8 AU for the sun), the FPB of an Earth-like planet is:

$$(F_p/F_*)_{IHZ} = 2.053 \times 10^{-10} / L_* \tag{15}$$

and

$$(F_p/F_*)_{OHZ} = 3.564 \times 10^{-11} / L_*. \tag{16}$$

This explains why, in general, A-type stars and giants are not ideal targets for direct-imaging of planets in reflected starlight; such stars are better suited for indirect detection of exoplanets via astrometry missions (Gould, Ford & Fischer 2003). Given a $10^{-10}$ limit on FPB, a sun-like star is actually too luminous for detection of an Earth at the outer HZ. Indeed, a $10^{-10}$ detection limit on fractional planet brightness implies that the maximum luminosity for stars whose planets are detectable at the outer HZ is only ~0.5 $L_\odot$ (~G8V), and habitable planets orbiting stars brighter than 2.4 $L_\odot$ (~F5V) are not detectable at all.

**Figure 7** shows how the IWA and FPB limit "tune" the observatory to a specific range of stellar spectral types. An IWA of 65 milliarcseconds and FPB limit of 4 x $10^{-11}$ (i.e., $m_p - m_* = 26$ mag) are shown in the figure, and there are 17 stars (8 K stars and 9 G stars) whose entire habitable zones fall within this detection space assuming a pole-on orientation. Another 122 stars have coverage over the Earth-equivalent insolation distance (EEID), including 32 F stars, 68 G stars, 21 K stars, and 1 M star (prox Cen). The effect of decreasing the IWA is to include additional late G- and K-type stars, while lowering the limiting FPB includes more early G- and F-type stars. Assuming the simple detection limits shown on Figure 7, Venus, Earth, Jupiter and Saturn are all detectable at a distance of 10 pc (shown as "V", "E", "J" and "S" on the plot), while Mars ("M") is likely beyond detectability depending on the exact profile of limiting fractional planet brightness. In reality, the NWO architecture will not have a perfectly sharp cutoff in FPB limit at the inner working angle, and planets which are brighter than the limiting fractional planet brightness should be detectable at slightly smaller angles.

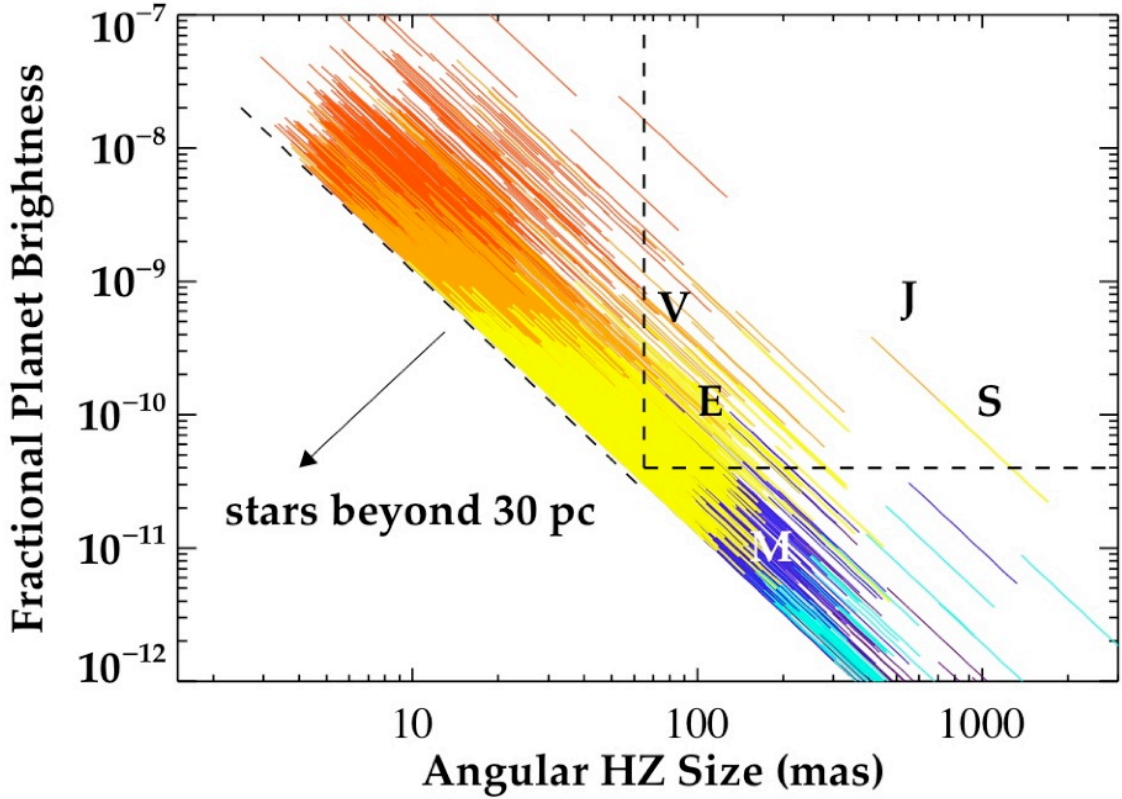

**Figure 7.** Fractional planet brightness as a function of angular separation, for planets seen in reflected starlight throughout the habitable zones of stars within 30 parsecs (color-coded by spectral type as in Figures 5 and 6), assuming the systems are seen pole-on (i.e., planets are seen at quadrature). Planets are Earth-like in size and albedo. The dashed lines show the NWO detection space on FPB limit = 4 x $10^{-11}$ and IWA = 65 milliarcsec. The Sun's planets are also plotted for a distance of 10 parsecs. As seen from a distance, Jupiter and Venus are the most detectable planets in the Solar System, each of them being several times brighter than the Earth.

### 2.4 Baseline IWA and Limiting FPB

As we have seen from Figure 7, any mission to directly image habitable planets will have a sample of targets that is biased toward particular stellar spectral types, depending on the IWA and FPB limits that are achieved. From Figure 7, it appears that the specifications IWA = 65 mas and limiting FPB = 4 x $10^{-11}$ (i.e., $m_p - m_* = 26$ magnitudes) are close to what will be required for a successful NWO, and these limits clearly tune the mission to sun-like (G-type) stars. To briefly illustrate the technology involved, an IWA of 65 mas combined with a limiting FPB of 26 magnitudes corresponds to a starshade that is 50-m in effective diameter at a distance of ~80,000 km (see discussion in §3). To maintain either of these performance parameters while improving the other (or to improve both) requires a larger starshade, operated farther away (Arenberg et al. 2008).

Based on the simple analysis above, we propose a baseline performance requirement for NWO of IWA = 65 mas, and limiting FPB = 4 x $10^{-11}$. In the following Sections, we test these performance requirements to determine whether they truly represent a mission that can achieve the objectives outlined in Table 1 given (1) the full range of possible planet orbits, (2) potentially high levels of exozodiacal light and low $\eta_\oplus$, and (3) a realistic target observing sequence.

### 2.5 Completeness

While Figure 7 provides a useful starting point in designing the NWO mission, these calculations simply assumed an Earth seen at quadrature phase. This does not take into account the fact that planetary systems will be seen at random inclinations and phases, or that planetary orbits are not necessarily circular.

To more accurately assess the numbers of target stars available with NWO and to provide a means of judging the relative science return from each star, we calculated the *completeness* for each star on the Hipparcos list (Brown 2005). The completeness is the probability that NWO would detect an HZ resident planet in a single visit, given the possible range of orbital positions and limitations of the observatory. Planets are considered to be detectable if they have an angular separation from the star greater than the IWA and a fractional brightness greater than the limiting FPB. In this paper, we assumed Earth-sized planets with a constant albedo (so that a planet's brightness around a given target varies only according to its distance from that star). Future work should take into account the variation of planet detectability for reasonable ranges in planet size and albedo.

For each star, we create a set of 900,000 hypothetical planet orbits that fall completely within the HZ. Each of these orbits is defined by five parameters: the semi-major axis, the eccentricity, the inclination of the orbital plane, the argument of periastron (angle between the ascending node, the point where the orbit crosses the plane of the sky, and the point of periastron), and the true anomaly (the angle between the direction of periastron and the position of the planet). The semi-major axes of the planet orbits are uniformly distributed, while the eccentricity at each semi-major axis is distributed uniformly within the ranges that allow for orbits to be entirely contained within the HZ (as in Arenberg & Schuman 2006; Arenberg, Knutson & Schuman 2005). The orientation of the orbit is uniformly distributed over the sphere, i.e., the cosine of the inclination angle is uniformly distributed within the range [-1, 1) and the argument of periastron is uniform in [0, 2π). Rather than randomly distributing the true anomaly, we chose 150 equally spaced time steps spanning one orbital period. From each "planet" in this ensemble, we calculated the angular separation and FPB. Finally, we determine what fraction of these planets NWO could detect, given a specific IWA and FPB limit.

The detectable fraction of habitable planets is the completeness for that star; this number also represents the probability of finding a HZ resident at that star in a single observation if every star has one habitable planet of Earthlike size and albedo (i.e., $\eta_\oplus$=1). As an example, **Figure 8** shows the full set of calculations for planets in the habitable zone of alpha Centauri B (completeness = 76%), where detectable planets are shown in green, given the IWA and FPB limit of 65 mas and 4 x $10^{-11}$, the baseline performance requirements proposed above. Orbits at the IHZ, EEID, and OHZ are shown with eccentricity = 0 and 60° inclination. The line corresponding to planets at quadrature is also shown (blue dashed line); this line corresponds to the HZ location previously plotted in Figures 6 and 7 (0.55-1.3 AU for alpha Cen B).

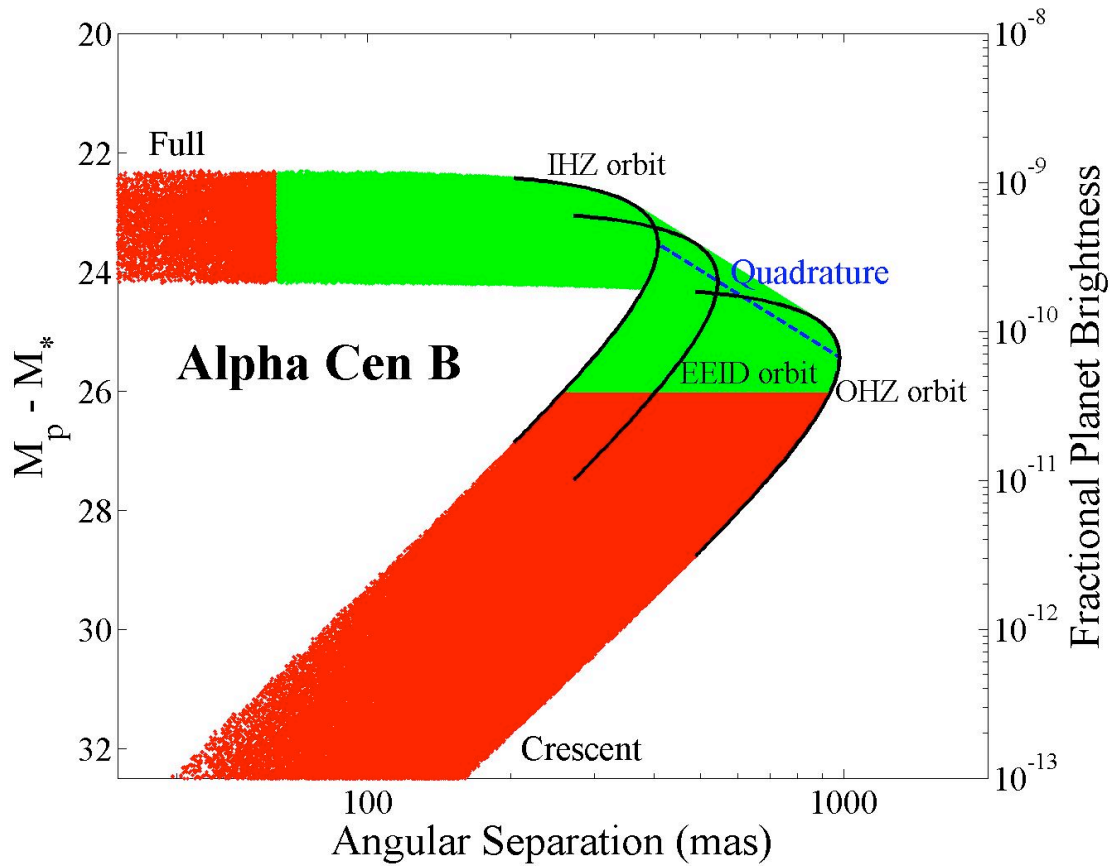

**Figure 8.** The fractional brightness and offset of all possible HZ residents for alpha Centauri B. The regions coded green represent planets detectable to NWO given an IWA of 65 mas and a fractional planet brightness limit of 4 x $10^{-11}$; the planets colored red would not be seen. In this case, 76% of the test planets can be seen, so the completeness for this star is 76%. Sample orbits are shown for planets at the IHZ, EEID, and OHZ (black curves). These orbits have eccentricity=0 and inclination=60°. The blue dashed line represents planets seen at quadrature and corresponds to the HZ size shown in Figures 6 and 7.

*Completeness and spectral type.* Given the baseline performance requirements proposed above (IWA = 65 mas and limiting FPB = 4 x 10⁻¹¹), there are 186 stars from the Hipparcos list with a completeness > 20%. **Figure 9** shows the distribution of these relatively high completeness targets by spectral type, and their numbers are given in **Table 3**. As should be expected from our discussion in the previous sections, the majority of stars with completeness > 20% are G-type, while the very nearest K-type stars (e.g., alpha Cen B) have the highest individual completeness values. As discussed above, the fainter K and M stars have a more favorable *fractional* planet brightness than G and F stars, and the *absolute* brightness of planets is roughly the same across all spectral types, but except for the very nearest systems, the habitable zones of K and M stars are lost within the IWA.

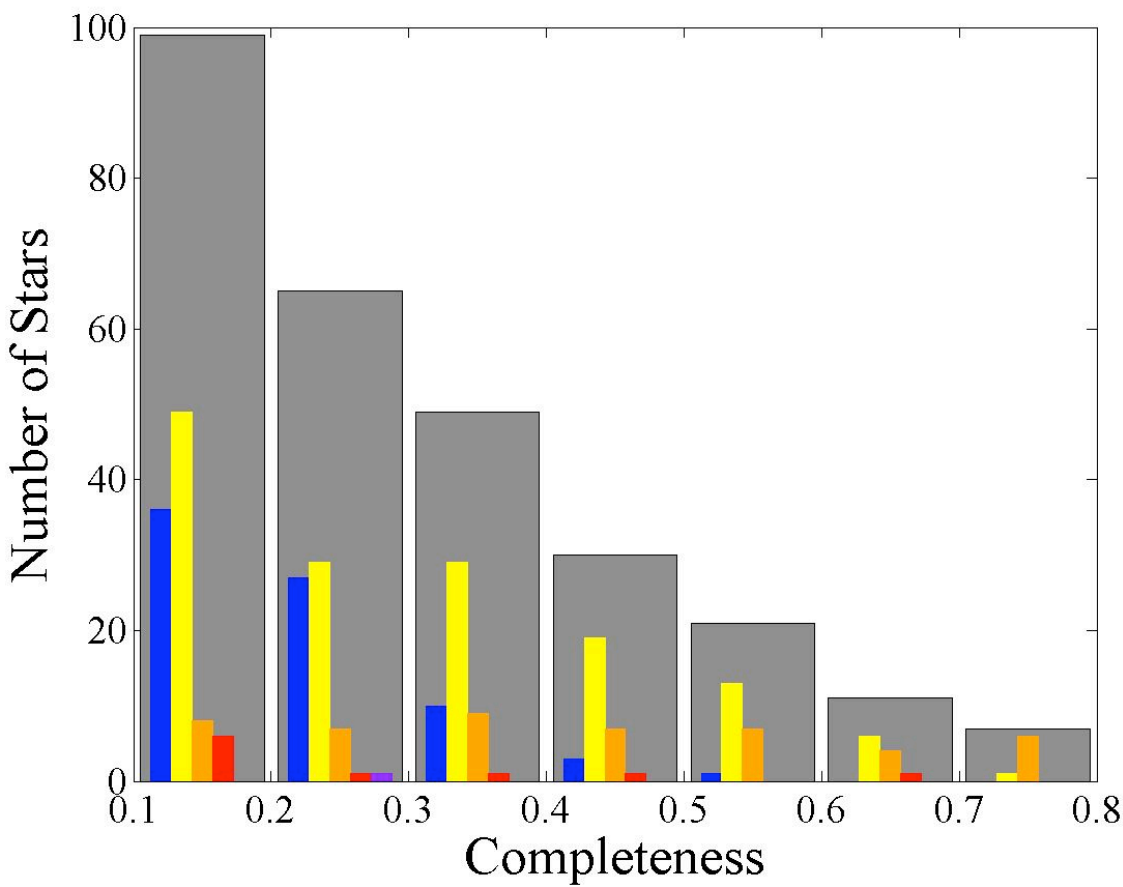

**Figure 9.** Histogram showing the number of stars versus completeness for a direct-imaging system like NWO. The criteria are FPB limit = 4 x 10⁻¹¹ and IWA = 65 mas. Within each bin we show the number of stars in that completeness range with a given spectral type, color-coded as in Figure 7: red = M dwarfs, orange = K, yellow = G, blue = F, cyan = A/B, purple = giants.

Table 3
Stars with Completeness > 20% for IWA=65 mas and $\Delta m_{limit}$=26 mag

| Spec Type | N | <Dist> (pc) | <V> (mag) |
|---|---|---|---|
| A/B | 0 | -- | -- |
| F | 41 | 16 | 5 |
| G | 97 | 14 | 5 |
| K | 43 | 8 | 6 |
| M | 4 | 5 | 9 |
| Giants | 1 | 9 | 4 |

**Figure 10** shows the completeness for each star in the sample versus (a) distance and (b) V-band magnitude. These plots show that the nearest F-, G- and K-type stars are favored, with very few viable targets fainter than V~7 or beyond 20 pc. A few M-type stars have significantly elevated completeness values due to their large bolometric corrections (which results in a larger calculated angular habitable zone size for a given V-band magnitude without decreasing the FPB). Given their highly favorable planet:star flux ratios and close proximity to the sun, these stars are worthy of some detailed consideration. Prox Cen is among the least luminous stars in the Hipparcos Catalogue, and only its proximity to the Sun allows us to consider it as an NWO target. In our calculations, a bolometric correction of -5.5 mag to proxima Centauri results in a completeness value of 62% despite an apparent magnitude of V=11 (not shown in Figure 10b). This is highly suspect, and more thorough study of proxima Cen's habitable zone location (and that of other M stars) is necessary to determine whether they are indeed appropriate targets for NWO.

It is also important to remember that completeness calculations reflect the *fraction* of the habitable zone searchable for planets, and this fraction can be high even though the physical search space is relatively small. In future studies, we might well ask: Is a 50% completeness M star (with an HZ annular width of ~0.15 AU) as desirable a target as a 50% completeness late F star whose HZ may be more than ten times wider (~1.5-2 AU)?

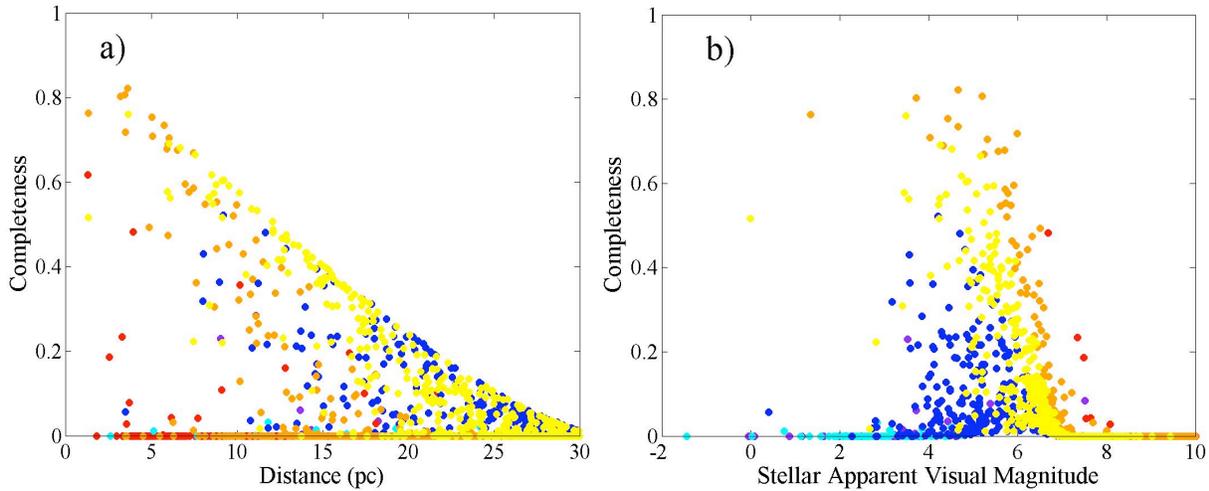

**Figure 10.** (a) Completeness versus distance and (b) completeness versus stellar V magnitude for all the stars in the Hipparcos 30 pc sample. The stars are color-coded by spectral type as in the previous Figures. The nearest K and G stars have the highest completeness values, while F stars dominate at larger distances. A few M stars, most notably proxima Cen in (a), appear elevated in completeness due to large bolometric corrections and correspondingly larger habitable zones; a more detailed study of their habitable zones locations is necessary to determine whether these are indeed viable targets.

### 2.6 *Exposure Time*

#### 2.6.1 *Effects of Exozodiacal Dust*

The integration times required to see the planets within the NWO detection space shown in Figure 7 depend strongly on the background present in the field of view, and this determines whether the scientific objectives of the mission can be achieved within a ΔV-limited, 5 year mission lifetime. Of particular concern is starlight scattered by dust surrounding the target star. Disks of tenuous dust surround even mature planetary systems like our own. In the Solar System, the zodiacal dust is produced by evaporation of comets and collisions between asteroids. Dust produced by similar processes is seen at much higher levels around many nearby main sequence stars. These systems, called debris disks (Gillett 1986), are optically thin and contain modest amounts of dust and sometimes small amounts of gas (e.g. Roberge et al. 2006). Even the densest ones have only a few lunar masses of dust (e.g. Dent et al. 2000). These disks are found around main sequence stars with ages ranging from about 10 Myr to several Gyr, although there is tentative evidence of debris disks as young as ~ 5 Myr (Currie et al. 2008, Hernandez et al. 2008). On average, the amount of dust declines with increasing stellar age (Su et al. 2006).

Observers generally quantify the amount of dust in a debris disk using the system's fractional infrared luminosity, $L_{IR}/L_*$, which is the starlight absorbed by the circumstellar dust and re-emitted at IR wavelengths relative to the stellar luminosity. The currently known debris disks have $L_{IR}/L_*$ values in the range $10^{-3} - 10^{-5}$ (e.g. Bryden et al. 2006). The zodiacal dust interior to our asteroid belt has $L_{IR}/L_* \approx 10^{-7}$ (Backman et al. 1993); this is below current detection levels for other stars. A system's fractional infrared luminosity is often given in units of "zodis," where one zodi of dust simply corresponds to $L_{IR}/L_* = 10^{-7}$ and is not a unit of dust mass or surface brightness. The fractional infrared luminosity is proportional to the debris dust mass, but other factors (e.g., grain properties) also affect $L_{IR}/L_*$ .

Debris disks often display dust structures, like rings (e.g. Schneider et al. 2009), clumps (e.g. Greaves et al. 2005), and warps (e.g. Krist et al. 2005). It's generally believed that such structures are produced by the dynamical influence of planets, similar to the clumps of zodiacal dust leading and trailing the Earth (Dermott et al. 1994). This explanation was confirmed for one of the best-studied debris disks, Fomalhaut, where high-contrast optical wavelength images showed that the circumstellar dust was confined to a narrow ring (Kalas et al. 2005). The morphology of this ring pointed to the presence of a giant planet, which was subsequently imaged near the expected position (Kalas et al. 2008).

While debris disk dust structures can provide indirect evidence of planets right now, they are also likely to be troublesome sources of confusion in future imaging of exo-Earths, as mentioned in §1.2.3. Dust clumps might be confused with an unresolved planet in broadband images, especially since dynamical clumps would orbit the star, though not necessarily with the same period as the perturbing planet (e.g. Kuchner & Holman 2003). The broadband color of the dust may not resolve the ambiguity, as debris disks are known to show diverse colors, from red to grey to blue (e.g. Golimowski et al. 2006, Krist et al. 2005). However, as aggregates of solid amorphous or crystalline material (e.g., pyroxenes, silicates, olivines, sulfides, amorphous carbon), interplanetary dust should have a relatively featureless reflectance spectrum at optical wavelengths (e.g., as seen in the reflectance spectra of interplanetary dust particles and meteorites; Bradley et al. 1996). This is easily distinguished from a habitable exoplanet with an atmosphere, which will show deep absorption features corresponding to gaseous species (e.g., $H_2O$, $O_2$, $CO_2$; Turnbull et al. 2006). Multi-band photometry may be an efficient way to distinguish dust clumps from true planets, and future work should explore medium- or narrow-band filter combinations that would be useful in this regard.

Meanwhile, the most certain effect of exozodiacal dust on direct observations of exo-Earths is to increase exposure times. Light scattered off the local zodiacal dust and the exozodiacal dust around nearby stars (exozodi, for short) will be mixed with the planet signal in both images and spectra. Even if nearby systems have exozodi levels no greater than the Solar System level, zodiacal and exozodiacal background will be the largest source of noise, assuming that light from the central star is suppressed by a factor of at least $10^{-10}$ relative to the star at the location of the habitable zone (Brown 2005).

2.6.2 *Detection Time*

Direct imaging of exoEarths in the habitable zones of nearby stars likely will be limited by background zodiacal and exozodiacal light. Following the approach in Brown (2005), the exposure time to image a planet in some bandpass is

$$t_{image} = \frac{2n_x\lambda^2}{\pi F_0 \Delta\lambda D^4 T}\left(\frac{S}{N}\right)^2 10^{0.8(M_p + 5\log d - 5)}\left[\left(\frac{206265''}{1\,rad}\right)^2\left(10^{-0.4z} + \varepsilon\,10^{-0.4x}\right) + \zeta\,10^{-0.4m_*}\left(\frac{\pi D^2}{4\lambda^2}\right)\right], \quad (17)$$

where $n_x$ = number of "noise pixels" in a critically sampled diffraction-limited image (calculated as the inverse of the sharpness, approximately 1/0.08 for diffraction limited imaging with a circular aperture; Burrows 2003), $\lambda$ = central wavelength of the image bandpass, $F_0$ = specific flux for zero magnitude in the bandpass, $\Delta\lambda$ = bandpass width, $D$ = diameter of the telescope aperture, $T$ = total facility throughput, $S/N$ = desired signal-to-noise of the planet detection, $M_p$ = absolute magnitude of the planet in the image bandpass, $d$ = distance to the system in parsecs, $z$ = surface brightness of the zodiacal dust in magnitudes arcsec$^{-2}$, $\varepsilon$ = exozodi surface brightness in units of the surface brightness of one zodi of exozodiacal dust, $x$ = surface brightness of one zodi of exozodiacal dust in magnitudes arcsec$^{-2}$, $\zeta$ = contrast level in the detection zone with respect to the theoretical peak of the stellar image, $m_*$ = apparent stellar magnitude in the image bandpass, and $\pi D^2/4\lambda$ = theoretical peak brightness of the stellar point-spread function. Typical values for these parameters appear in **Table 4**. We have not included dark noise and read noise in these calculations, both of which are negligible compared to shot noise assuming exozodi levels at least as high as that in our own Solar System (Brown 2005).

Table 4
Typical values for the parameters in the imaging exposure time equation

| Parameter | Typical Value | Explanation |
|---|---|---|
| $n_x$ | 1/0.08 = 12.5 pixels | Number of pixels in a critically sampled diffraction-limited image |
| $\lambda$ | 600 nm | Central wavelength of the image bandpass |
| $F_0$ | 9500 photons sec$^{-1}$ cm$^{-2}$ nm$^{-1}$ | Specific flux for zero mag in $V$ band |
| $\Delta\lambda$ | 200 nm | Bandpass width |
| $D$ | 400 cm | Telescope aperture diameter |
| $T$ | 0.50 | Estimated total throughput for NWO |
| $S/N$ | 10 | Signal-to-noise |
| $M_p$ | 29.68 | Absolute $V$ magnitude of the Earth at quadrature |
| $z$ | 23 mag arcsec$^{-2}$ | Surface brightness of the zodiacal dust |
| $x$ | 22 mag arcsec$^{-2}$ | Exozodi surface brightness for a Solar System twin (1 zodi) viewed at 60° inclination. Calculated with the ZODIPIC code (2006 TPF-C STDT Report). |
| $\zeta$ | $3\times10^{-11}$ | Contrast level in the detection zone relative to the theoretical peak of the stellar PSF |
| $m_*$ | 4.48 | $V$ magnitude of the Sun at 10 parsec |

In Equation 17, the first term in the square brackets accounts for the local zodiacal background, the second accounts for the exozodi background, and the third for the residual unsuppressed stellar light. For our purposes, a "Solar System twin" has 1 zodi of exozodiacal dust with properties identical to those of the zodiacal dust.

The contrast level in the detection zone ($\zeta$) is an estimate of the amount of residual starlight diffracted by the starshade, using the current NWO baseline design. This value must be low enough to allow the detection of the exoplanet against this background light, both from a noise and a speckle-confusion standpoint. The exact relationship between the stellar contrast (i.e., the surface brightness of starlight in the detection zone with respect to the theoretical Airy peak of the stellar image), $\zeta$, and the faintest detectable planet (i.e. the limiting FPB) is still uncertain, but preliminary analyses indicate that $\zeta$ must be less than or approximately equal to the limiting FPB. In the following exposure time calculations, we assume $\zeta = 3 \times 10^{-11}$.

As mentioned in §2.3, the NWO architecture will not have a perfectly sharp cutoff in FPB limit at the inner working angle. Glassman et al. (2010) have described the wavelength dependence of the starlight suppression, and they defined the IWA to be the angle where the starshade throughput drops to 50%. For the purposes of this paper we treat the IWA as a simple step function in starshade throughput that drops from 100% outside the IWA, to $3 \times 10^{-11}$ within the IWA. Outside of the IWA, the total facility (telescope, instruments, and starshade) throughput is taken to be $T = 50\%$ in the detection bandpass.

The telescope aperture diameter strongly affects the imaging exposure time, with a factor of $D^{-2}$ for the collecting area and another factor of $D^{-2}$ for the telescope spatial resolution, since a smaller PSF adds less zodiacal and exozodi background to the planet signal. As a useful benchmark, the NWO baseline design can achieve $S/N = 10$ on an Earth-like planet in a Solar System twin at 10 pc in about 2 hours.

**Figure 11** shows a plot of exposure time versus exozodi brightness ($\varepsilon$ in Equation 17) for an Earth-Sun twin at distances between 5 and 15 pc, using the parameter values given in Table 4. The lines plotted give the times to get $S/N = 10$ on Earth-like planets at 1 AU from Sun-like stars at different distances. As can be seen, the exposure time increases (1) linearly with exozodi brightness and (2) more steeply with increasing distance. Exposure times for more distant stars are strongly sensitive to exozodi brightness, because the physical area encompassed by the PSF (and the amount of exozodi background flux contained in it) grows with increasing distance.

While Figure 11 serves to show the general dependence of exposure time on exozodi brightness, **Figure 12** shows exposure times for imaging Earth-like planets orbiting specific stars given the baseline NWO mission (IWA= 65 mas, limiting FPB = $4 \times 10^{-11}$). For each system, the planet was assumed to be at the EEID (1 AU for a solar twin) and its magnitude was calculated using the formulae in §§2.1-2.3. The exozodi level at this location is assumed to be a constant for all stars, and we made no attempt to account for clumps or for declining exozodi brightness at larger separations from the star. Each star is plotted as a bar showing the imaging exposure times for exozodi surface brightnesses between $\varepsilon = 1$ (brightness of one zodi of exozodi) and $\varepsilon = 100$ (100 times brighter than one zodi of exozodi). The values of the other parameters in Equation 17 were taken from Table 4. Among stars with non-zero completeness for the baseline NWO, there are 266 and 77 targets with exposure times less than 1 day for $\varepsilon = 1$ and $\varepsilon = 10$, respectively. Even at $\varepsilon = 100$, there are 66 targets with exposure times less than one week. Figure 12 shows that exposure times increase with decreasing stellar luminosity, although G- and K-type stars are still the best targets for NWO due to their proximity to the sun. Beyond ~20 pc distance, no targets have a detection time of less than 1 day.

One striking feature of Figure 12 is the large range in exposure time for proxima Centauri, the nearest M dwarf. As mentioned in §2.5, this is due to the high bolometric correction applied to very red stars. If this correction accurately locates the habitable zone for proxima Cen, then the relatively high completeness and favorable fractional planet brightness of this star might present a special opportunity for NWO. However, as Figure 12 shows, the presence of any exozodiacal dust around prox Cen might render planets undetectable within a reasonable exposure time.

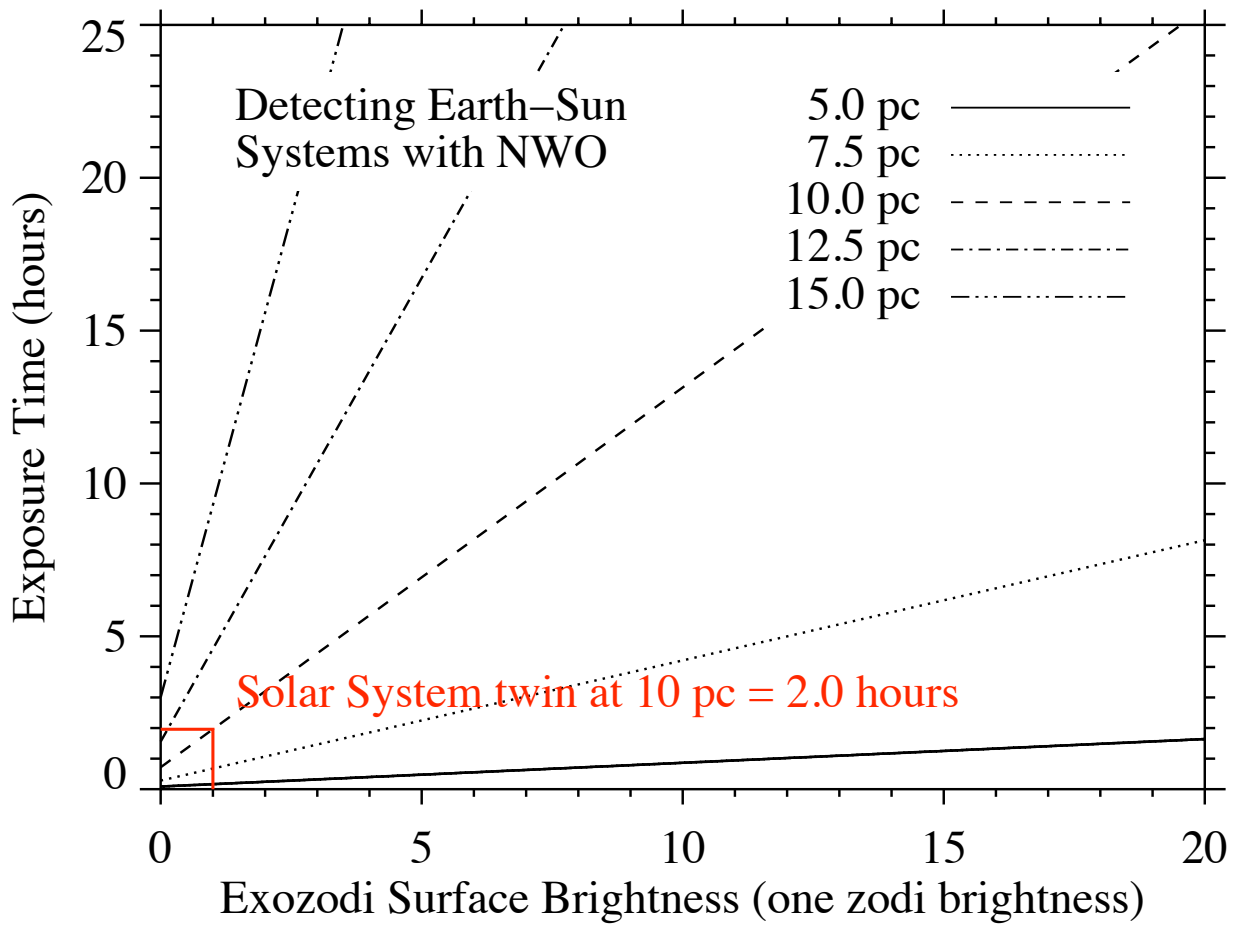

**Figure 11.** Planet imaging exposure time versus exozodi brightness. The y-axis shows the time to detect an Earth-like planet in the habitable zone at $S/N = 10$, calculated using Equation 17 and the parameter values in Table 4. The x-axis shows the exozodi surface brightness, in units of one zodi. The exozodi surface brightness is proportional to $10^{-0.4x}$, where $x = 22$ mag arcsec$^{-2}$ for a system viewed at 60° inclination.

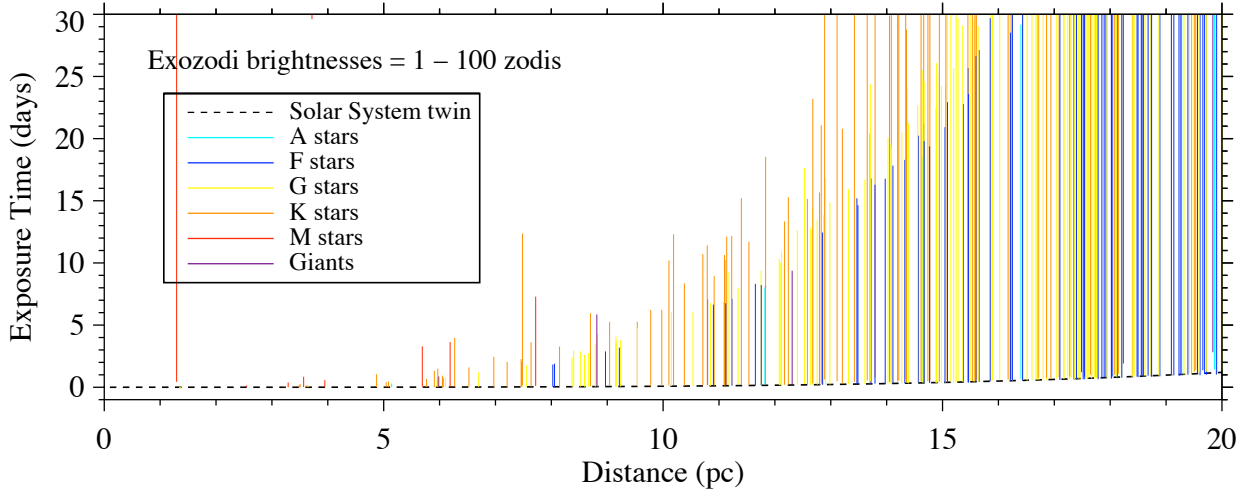

**Figure 12.** Exo-Earth imaging exposure times versus distance for Hipparcos stars. All times are to achieve $S/N = 10$ on an Earth-like planet orbiting at the EEID. Each star is plotted as a bar showing the times for exozodi brightness levels between $\varepsilon = 1$ and $\varepsilon = 100$, with spectral types indicated by color. The dashed line shows the time to image the Earth in a Solar System twin at that distance.

### 2.6.3 *Characterization Time*

How much additional time is required to characterize the planet candidates detected by NWO? Levine, Shaklan & Kasting (2006) discussed the required spectral resolution for various optical and near-IR spectral features due to oxygen, ozone, $CO_2$, water and methane, and a spectral resolution $R = \lambda/\Delta\lambda_{spec} = 100$ is adequate to detect these. The integration time required for NWO to take a spectrum with resolution $R = \lambda/\Delta\lambda_{spec} = 100$ and $S/N = 10$ is approximately

$$t_{spec} \approx \left(\frac{\Delta\lambda_{image}}{\Delta\lambda_{spec}}\right) \times t_{image} = \left(\frac{\Delta\lambda_{image}}{\lambda/R}\right) \times t_{image} = \left(\frac{200\,\text{nm}}{600\,\text{nm}/100}\right) \times t_{image} = 33 \times t_{image} \qquad (18)$$

where $\Delta\lambda_{image}$ is the imaging bandpass, $\Delta\lambda_{spec}$ is the width of one resolution element in the spectrum, $t_{image}$ is the imaging exposure time for $S/N = 10$, $\lambda$ is the central wavelength of the spectrum, and $R$ is the spectral resolution. Given simulations showing that the mean time to move the starshade between targets is 11 days (see §2.7.4), we are unlikely to spend less than 1 day imaging a system. With a minimum imaging time of 1 day, a blind application of Equation 18 would give a minimum spectroscopy time of 33 days. However, stars with imaging times < 1 day do not require this much time to get $S/N = 10$ spectra. There will be some minimum spectrum $S/N$ and exposure time required to decide whether the candidate is a real planet or not. For now, we set the minimum spectroscopy time to be 1 day and require spectra with $S/N \geq 10$ for all planet candidates.

Ignoring for the moment the possibility of confusion from exozodi clumps and background sources, the total time spent on spectroscopy will be approximately $\eta_\oplus$ times the total time for doing spectroscopy at each target, times the completeness of each target, $C_n$, summed over all targets acquired. Therefore, the total integration time on $n$ target stars is

$$t_{\text{int}} = \sum_{n \text{ stars}} \left( t_{image,\,n} + t_{spec,n} \right) \approx \sum_{n \text{ stars}} \left( t_{image,n} + 33.3\,\eta_{\oplus} C_n t_{image,n} \right) . \tag{19}$$

Acquiring each target takes on average 11 days of starshade travel time (see §2.7.4).  So the total mission time is

$$t_{\text{int}} = \sum_{n \text{ stars}} \left( t_{image,\,n} + t_{spec,n} + t_{travel,n} \right) \approx \sum_{n \text{ stars}} \left( t_{image,n} + 33.3\,\eta_{\oplus} C_n t_{image,n} + 11\,\text{days} \right) . \tag{20}$$

These times for planet imaging, characterization, and movement of the starshade will guide our mission performance analysis in the following Sections.

## 2.7 *Mission Performance*

The key performance metric for any exo-Earth imaging mission, regardless of architecture, is the total number of Habitable Zones that can be searched during the mission lifetime.  In this Section, we prioritize the 30 pc sample of potential targets according to observational efficiency and scientific merit.  We then use this prioritization to find out how many targets must be observed within the mission lifetime in order to achieve a 95% probability of detecting at least one habitable planet for any mission with IWA= 65 mas, and limiting FPB = 4 x 10^{-11}.  To determine whether these targets can indeed be observed within the NWO mission lifetime, we run (1) a simple analytical code and (2) a more realistic Monte Carlo scheduling simulator to choose targets and predict the (IWA= 65 mas, limiting FPB = 4 x 10^{-11}) baseline mission performance within the mission lifetime, as a function of exozodiacal brightness levels and prevalence of exo-Earths.

### 2.7.1 *Prioritizing the Search Targets*
In **Figure 13**, the NWO imaging exposure times from Figure 12 are plotted versus the completeness values calculated in Section 2.5.  As in Figure 12, each star is plotted as a bar showing the exposure times for exozodi levels between $\varepsilon = 1$ and $\varepsilon = 100$, with different colors for different types of stars.

For reference, the shaded area of Figure 13 shows stars of relatively high completeness and low exposure time.  This area includes 173 stars with detection times $\leq 1$ day for $\varepsilon = 1$ and completeness greater than 20%.  These stars have a cumulative completeness of 69 total HZs.  Summing over all 672 stars with nonzero completeness for the baseline mission, we find that 99 total HZs are potentially available for study with the baseline detection limits.

The calculated completeness values and exposure times can be used to prioritize targets according to scientific merit and observational efficiency.  To identify the stars most favorable for observation with NWO, we have calculated a figure of merit for each system based on both completeness and required integration time.  We assume that all stars have the same exozodi brightness, and this level is varied from $\varepsilon = 1$ to $\varepsilon = 100$.  We then calculated a figure of merit, $F_{merit}$, for each star at each exozodi brightness level:

$$F_{merit,\,n} = C_n / t_{image} . \tag{21}$$

By this measure, the well-known stars ε Indi, ε Eri, 61 Cyg A&B, α Cen A&B, τ Ceti, 40 Eri A, 70 Oph, and σ Dra constitute the ten highest priority stars assuming a "medium" exozodi level of ε = 10. However, not all of these stars are necessarily ideal targets, as shall be discussed in Paper 2. In the following sections, $F_{merit}$ is used as a weight in selecting stars for observations and determining mission performance.

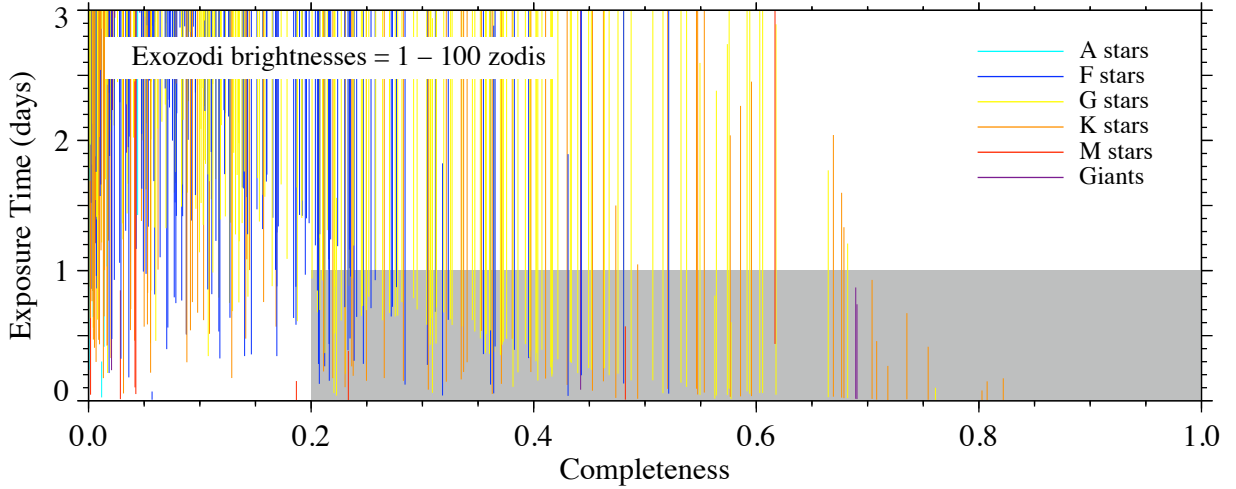

**Figure 13.** Exposure time for planet detection, versus completeness for the NWO baseline mission. All times are to achieve $S/N = 10$ on an Earth-like planet orbiting at the EEID. Each star is plotted as a bar showing the times for exozodi brightness levels between ε = 1 and ε = 100, with spectral type indicated by color. For reference, the shading shows $t \leq 1$ day and completeness > 20%, including 173 stars and 69 total HZs at ε = 1.

### 2.7.2 *Required Number of Targets*

As outlined in Table 1, a successful mission must achieve a 95% probability of detecting at least one Earthlike planet, assuming that $\eta_\oplus \geq 0.1$ and that the background of exozodiacal light for our targets is up to 100 times that of our own system. In §1.2 we discussed the probability of detecting at least one planet in the case of $n$ target stars observed at 100% completeness. To calculate the probability of detecting at least one planet in the case of $n$ target stars with various completeness values $C_n$, we note that after observing one star, the probability that a planet will have been found is $\eta_\oplus C_1$. Therefore the probability that a planet will *not* be found after observing the first target star is simply:

$$P(N_{php} = 0,\, n = 1) = 1 - \eta_\oplus C_1, \tag{22}$$

where $N_{php}$ is the number of potentially habitable planets found, and $n$ is the number of targets searched.

Observing a second target, the probability that zero planets were found drops to:

$$P(0,2) = P(0,1)(1 - \eta_\oplus C_2) = (1 - \eta_\oplus C_1)(1 - \eta_\oplus C_2), \tag{23}$$

and after observing the third target the probability of detecting no planets becomes:

$$P(0,2) = P(0,1)(1 - \eta_\oplus C_2) = (1 - \eta_\oplus C_1)(1 - \eta_\oplus C_2)(1 - \eta_\oplus C_3), \qquad (24)$$

and so on, such that the probability of detecting no planets in a program of $n$ target stars, where the completeness of each system is $C_n$ is:

$$P(0,n) = P(0, n-1)(1 - \eta_\oplus C_n). \qquad (25)$$

Therefore, the cumulative probability of detecting at least one planet after observing $n$ target stars is

$$P(N_{php} > 0, n) = 1 - P(0,n) = 1 - (1 - \eta_\oplus C_n)[1 - P(>0, n-1)]. \qquad (26)$$

Equation (26) gives us a recursive method for finding P($N_{php} > 0$, $n$), assuming various values of $\eta_\oplus$ and given completeness calculations for the individual targets observed with our specified detection limits. To do this, we sorted the stars in our 30 parsec Hipparcos database by decreasing $F_{merit}$, simulating an observation program where the highest priority stars are observed first. Starting with our highest priority star at each exozodi level, we used the recursion formula (26) to calculate the cumulative probability of detecting at least one planet after each observation, for various values of $\eta_\oplus$. **Figure 14** shows, for ε = 10, how the probability of detecting at least one planet climbs with successive observations for various values between 0.01 ≤ $\eta_\oplus$ ≤ 1. The probability climbs rapidly at first as stars with the highest completeness and shortest exposure times are observed early on, gradually giving way to stars with lower completeness but still relatively short exposure times. **Table 5** lists, for various levels of exozodi and $\eta_\oplus$, the number of targets, $N_{targets}$, that must be observed with the baseline (IWA= 65 mas, limiting FPB = 4 x $10^{-11}$) mission to achieve P($>0$, $N_{targets}$) = 95%, thus satisfying the first science objective. We also show the average completeness $\overline{C_n}$ of the targets observed, the total number of HZs searched, $\sum_n C_n$, (i.e., the cumulative completeness) in reaching this confidence level for each case, and the expectation value $\langle N_{php} \rangle$ for the number of potentially habitable planets found. The numbers shown in Figure 14 and Table 5 apply to any mission architecture.

Table 5 shows that the level of exozodiacal light has little impact on the number of targets required for a successful mission, reflecting the fact that the stars maintain nearly their same relative priority regardless of ε. More distant stars are more strongly affected by changes in exozodi brightness, but this results in only a few minor rearrangements in the target list when ordered by decreasing $F_{merit}$. We note that it is not strictly necessary to observe only the highest priority stars (or to observe them in order of descending priority). Indeed, in a real mission this may not always be possible due to solar avoidance and other constraints. A larger number of lower completeness stars may also be observed, as long as the cumulative completeness shown in Table 5 is achieved within the starshade ΔV limits and mission lifetime.

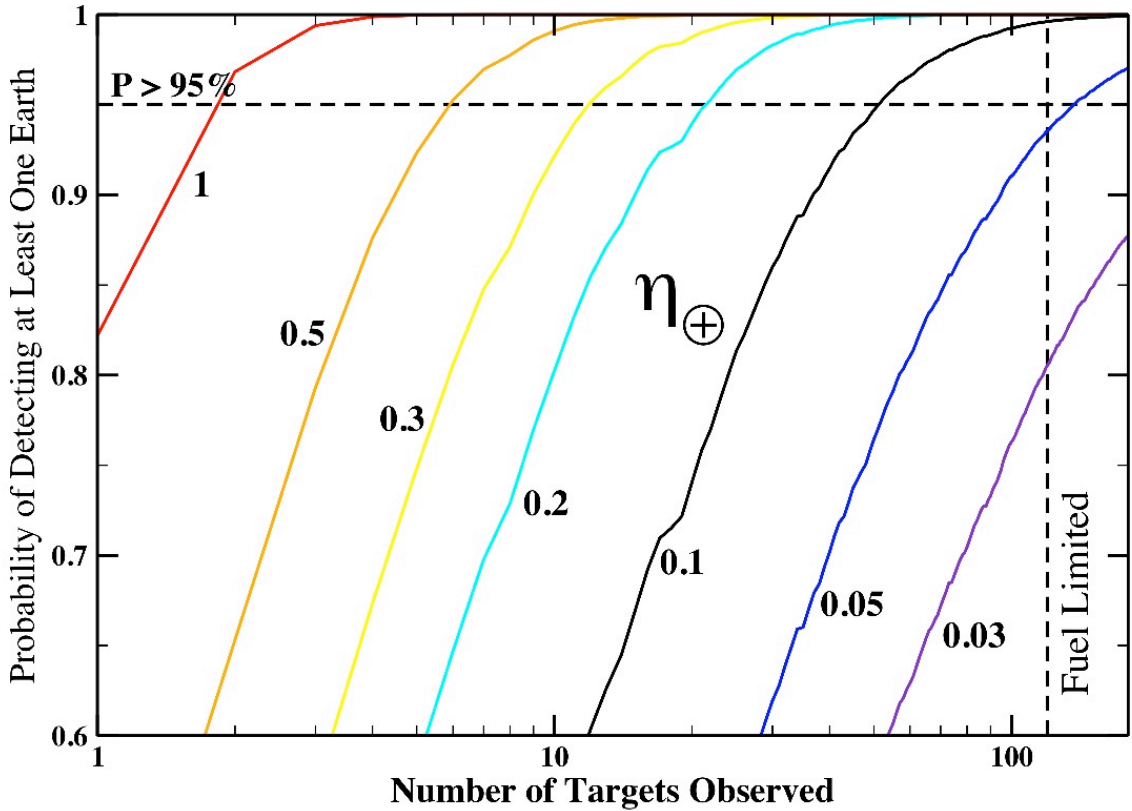

**Figure 14.** The probability of detecting at least one habitable planet as a function of the number of target stars searched, for a variety of $\eta_\oplus$ values and assuming the baseline detection limits (IWA = 65 mas, limiting FPB = $4\times10^{-11}$). All targets are assumed to contain exozodiacal light levels ten times that of the Solar System ($\varepsilon = 10$).

We note that the science objectives could also be met with slightly fewer stars (only ~47 instead of ~52 in the case of $\eta_\oplus$=0.1), if they are observed in order of descending completeness value instead of $F_{merit}$. Indeed, the curves in Figure 14 show occasional horizontal "steps", where relatively low completeness stars are ranked with high priority due to their very short exposure times; the suitability of these targets will depend on their location on the sky and the starshade travel time to reach them (discussed in §2.7.4).

In the case of $\eta_\oplus = 0.1$, we find that ~52 stars must be observed; this yields a cumulative completeness of 29 habitable zones and a corresponding expectation value in the number of potentially habitable planets detected $\langle N_{php} \rangle \sim 3$. This is in agreement with the simple example discussed in §1.2, and this result applies to all planet-imaging architectures. This analysis also reveals that if the fraction of stellar systems with Earthlike planets is very low, $\eta_\oplus < 0.03$, it is simply not possible to reach 95% probability of detecting at least one planet with these baseline detection limits (IWA = 65 mas, limiting FPB = $4\times10^{-11}$), because the total completeness of *all* target stars is not high enough.

Table 5
Required Number of Targets for 95% Probability to Detect at Least One Planet; Baseline NWO

| **Low Exozodi Brightness ($\varepsilon = 1$)** | | | | | | | | |
|---|---|---|---|---|---|---|---|---|
| $\eta_\oplus$ | 0.03 | 0.05 | 0.06 | 0.08 | 0.1 | 0.2 | 0.3 | 0.5 | 1 |
| $N_{targets}$ | 658 | 136 | 104 | 68 | 50 | 21 | 12 | 6 | 2 |
| $\overline{C_n}$ | 0.13 | 0.43 | 0.47 | 0.53 | 0.57 | 0.70 | 0.75 | 0.79 | 0.82 |
| $\sum_n C_n$ | 98 | 59 | 49 | 37 | 29 | 15 | 9 | 5 | 1.6 |
| $\langle N_{php} \rangle$ | 3 | 2.9 | 2.9 | 2.9 | 2.9 | 2.9 | 2.7 | 2.4 | 1.6 |
| $t_{det}$ (d) | 8300 | 136 | 104 | 68 | 50 | 21 | 12 | 6 | 2 |
| $t_{spec}$ (d) | 4800 | 31 | 18 | 7 | 4 | 3 | 3 | 3 | 2 |
| $t_{travel}$ (d) | 7238 | 1496 | 1144 | 748 | 550 | 231 | 132 | 66 | 22 |
| $t_{total}$ (yr) | 56 | 4.7 | 3.5 | 2.3 | 1.7 | 0.7 | 0.4 | 0.2 | 0.1 |

| **Medium Exozodi Brightness ($\varepsilon = 10$)** | | | | | | | | |
|---|---|---|---|---|---|---|---|---|
| $\eta_\oplus$ | 0.03 | 0.05 | 0.06 | 0.08 | 0.1 | 0.2 | 0.3 | 0.5 | 1 |
| $N_{targets}$ | 658 | 137 | 105 | 70 | 52 | 22 | 12 | 6 | 2 |
| $\overline{C_n}$ | 0.15 | 0.43 | 0.46 | 0.52 | 0.56 | 0.64 | 0.72 | 0.78 | 0.81 |
| $\sum_n C_n$ | 98 | 59 | 49 | 36 | 29 | 14 | 9 | 5 | 1.6 |
| $\langle N_{php} \rangle$ | 3 | 2.9 | 2.9 | 2.9 | 2.9 | 2.8 | 2.6 | 2.3 | 1.6 |
| $t_{det}$ (d) | 6e5 | 214 | 129 | 71 | 51 | 22 | 12 | 6 | 2 |
| $t_{spec}$ (d) | 4e5 | 210 | 116 | 44 | 20 | 3 | 3 | 3 | 2 |
| $t_{travel}$ (d) | 7238 | 1507 | 1155 | 770 | 572 | 242 | 132 | 66 | 22 |
| $t_{total}$ (yr) | 283 | 5.3 | 3.8 | 2.4 | 1.8 | 0.7 | 0.4 | 0.2 | 0.1 |

| **High Exozodi Brightness ($\varepsilon = 100$)** | | | | | | | | |
|---|---|---|---|---|---|---|---|---|
| $\eta_\oplus$ | 0.03 | 0.05 | 0.06 | 0.08 | 0.1 | 0.2 | 0.3 | 0.5 | 1 |
| $N_{targets}$ | 672 | 138 | 105 | 71 | 52 | 23 | 13 | 7 | 3 |
| $\overline{C_n}$ | 0.15 | 0.43 | 0.46 | 0.51 | 0.56 | 0.61 | 0.67 | 0.74 | 0.69 |
| $\sum_n C_n$ | 99 | 59 | 49 | 36 | 29 | 14 | 9 | 5 | 2 |
| $\langle N_{php} \rangle$ | 3 | 3 | 2.9 | 2.9 | 2.9 | 2.8 | 2.8 | 2.6 | 2.1 |
| $t_{det}$ (d) | 1.5e6 | 1660 | 808 | 275 | 122 | 24 | 13 | 7 | 3 |
| $t_{spec}$ (d) | 9e5 | 2048 | 1108 | 413 | 198 | 27 | 9 | 3 | 3 |
| $t_{travel}$ (d) | 7282 | 1518 | 1155 | 781 | 572 | 253 | 143 | 77 | 33 |
| $t_{total}$ (yr) | 6700 | 14.3 | 8.4 | 4 | 2.4 | 0.8 | 0.5 | 0.2 | 0.1 |

Table 5 also shows the integration times for detection ($t_{det}$) and characterization ($t_{spec}$), the NWO starshade travel time ($t_{travel}$), and the total mission lifetime to carry out observations of all $N_{targets}$ in each astrophysical scenario. We note that $t_{travel}$ changes little with increasing exozodi (since the same number of targets are required), but integration times are strongly affected at low $\eta_\oplus$. This analysis shows that there is no feasible "single visit" scenario in which the baseline NWO could achieve the science objectives if $\eta_\oplus \leq 0.06$ (shaded area in Table 5). If Kepler results indicate that $\eta_\oplus$ is smaller than ~6% among sun-like stars, a higher performance baseline design (i.e., larger starshade, as discussed in §3) should be considered.

*The Effect of Allowing Return Visits.* An important caveat to the above is that these results apply to the case where each star is visited only once. In that scenario, the optimal observing strategy is simply to start with the highest priority target and work our way down the target list to lower and lower priority stars. However, Arenberg, Knutson and Schuman (2005) described a strategy where, in the case of no initial planet detections, return visits could be made to the highest completeness targets. In that scenario, additional completeness can be gained, thereby raising the probability of detecting an Earth-like planet and allowing for a successful mission even in the case of $\eta_\oplus < 0.06$. However, the additional completeness that can be gained from a return visit is a function of time since the initial observation, and subsequent observations must be timed to maximize the probability of planet detection. The present analysis does not include the added complication of allowing return visits, but future work should develop a revisit strategy to take full advantage of the search space around the highest priority target stars.

*The Effect of Changing the Detection Limits.* What gains or losses occur with a different TPF or NWO performance? For the worst case scenario of $\eta_\oplus = 0.1$ and $\varepsilon = 100$, as specified by the science requirements, we repeated the above calculations for an "enhanced" mission, assuming either a smaller inner working angle (50 mas), a deeper limiting fractional planet brightness (dM = 27 mag), or both. We then considered lower performance "lite" designs with either a larger IWA (80 mas), a shallower limiting FPB (dM= 25 mag), or both. Finally, we considered two "alternative" designs where (1) the IWA was larger (80 mas) but limiting FPB was deeper (dM = 27 mag), and (2) the IWA was smaller (50 mas) but limiting FPB was shallower (dM = 25 mag).

**Table 6** shows, for the case of $\eta_\oplus = 0.1$ and $\varepsilon = 100$, how the mission program (in terms of the target stars, telescope time, and starshade travel time) is affected by the detection limits for a scenario where only single visits are allowed. For IWA = (50, 65 and 80 mas) and limiting FPB dM = (25, 26, and 27 mag), the Table shows (1) the number of targets that must be observed in order to achieve the "$P_\oplus \geq 95\%$" requirement, (2) the spread of spectral types included in the minimum single visit observing program, along with their cumulative completenesses and the telescope time, $t_{det}$, required to search their HZs for Earths, (3) the distance of the farthest target, (4) the total telescope time spent characterizing planets, $t_{spec}$, if all the shown program stars are searched, and (5) the minimum planet-finding mission lifetime, $t_{tot}$, including starshade travel time ($t_{travel}$, assuming an 11 day slew time between targets).

*NWO-"Lite" Concepts.* Relaxing the limiting FPB from 26 to 25 magnitudes significantly degrades the completeness of high priority stars, such that many more targets must be observed in order to achieve the "95%" criterion. Considering the starshade travel time, this would be very difficult to accomplish within a 5-year mission lifetime. On the other hand, relaxing the IWA from 65 mas to 80 mas is not necessarily fatal to the mission. The total integration time required is increased, but more than 50% of the telescope time is still available for other observations and the total mission time still falls within the 5 year limit. Relaxing both the IWA and the FPB limit, to 80 mas and 25 mag respectively, we find that it is not possible to achieve 95% likelihood of detecting a planet regardless of number of target stars. Only ~90 stars have non-zero completeness in this scenario, and observing all of them yields a probability of exo-Earth detection of only ~88%.

*"Tuning" NWO to Sun-like stars.* In terms of the spectral types observed, while the top targets for all of the NWO/TPF versions shown in Table 6 are heavy on K stars, the detection limits do have an impact on the spectral types of the highest priority stars. Earlier type stars become viable targets with deeper FPB limits and larger IWA (e.g., the case of IWA=80 mas, dM=27 mag where 11% of the completeness comes from seven F and A stars), while later type stars are favored with smaller IWA and higher FPB limits (e.g., the case of IWA=50 mas, dM=25 mag where >60% of the completeness comes from K stars, and only about 1% of the completeness comes from two F stars). The scenarios where G-type stars account for the bulk of the HZs searched are for the case of a larger IWA (80 mas) and a baseline or deeper FPB limit (dM=26, 27 mag).

However, because we have defined priority to be inversely proportional to exposure time, this "tuning" effect is not as noticeable as one would expect if targets were prioritized only according to completeness (as per our discussion in §2.3). By lowering the FPB limit, earlier type stars do gain in completeness, but due to the fact that all habitable planets have the same absolute magnitude, the nearest systems (i.e., late types) are still strongly favored. Indeed, for all of the versions of TPF/NWO in Table 6, the vast majority of the highest priority targets are located within 10 pc. Nearly three quarters of the stars in the Hipparcos Catalog within 10 pc are K- and M-type stars.

This situation could be changed by overriding the priority ranking and intentionally selecting sun-like and earlier type stars as long as the total completeness reaches the required level (~29 HZs for $\eta_\oplus$ = 0.1) within the constraints. Such a program would necessarily have a lower observing efficiency in terms of HZs/day, but may be more satisfactory in terms of finding Earth-sun analog systems. Given that only 52 stars and 320 days of integration time are needed to fulfill the completeness requirement for the baseline NWO, there is clearly some room for flexibility in constructing a target list that represents an appealing mix of spectral types (or other stellar characteristics), while still remaining within the starshade $\Delta V$ budget and mission lifetime. Indeed, the missions in Table 6 all have room to be adjusted in this manner, except for the two shaded areas where integration time already exceeds 5 years.

As an example of this, we sorted the 150 highest priority stars for the baseline mission according to B-V color and repeated the above recursive calculations for only those stars having B-V ≤ 0.9. From this emerged a list of 52 targets (i.e., the same number as before) that satisfy the "95% requirement", comprised of 1 A star, 9 F stars, 28 G stars and 14 early K stars. In this case, 56% of the HZs searched belong to G stars. Assuming high levels of exozodi ($\varepsilon$ = 100), this program would take 3.2 years total (including starshade travel time) to carry out, and requires only 1.6 years of telescope time.

Table 6.
Single Visit Programs w/95% Probability of Detecting at Least One Earth: $\eta_\oplus = 0.1$, $\varepsilon = 100$

**IWA = 80 mas**

| | dM = 25 mag | | | dM = 26 mag | | | dM = 27 mag | | |
|---|---|---|---|---|---|---|---|---|---|
| | $N_{targets}$ | $\sum_n C_n$ | $t_{det}$ (d) | $N_{targets}$ | $\sum_n C_n$ | $t_{det}$ (d) | $N_{targets}$ | $\sum_n C_n$ | $t_{det}$ (d) |
| A stars | 0 | 0 | 0 | 1 | 0.01 | 1 | 2 | 0.2 | 2 |
| F stars | 26 | 0.7 | 397 | 10 | 3.2 | 52 | 5 | 2.9 | 11 |
| G stars | 61 | 7.4 | 882 | 27 | 12.9 | 124 | 20 | 13.4 | 54 |
| K stars | 48 | 11.2 | 430 | 26 | 12.4 | 64 | 20 | 11.9 | 33 |
| M stars | 5 | 0.7 | 3e5 | 3 | 0.3 | 3 | 3 | 0.3 | 3 |
| TOTALS | 140 | <29 | >>5 yr | 67 | 29 | 243 | 50 | 29 | 103 |
| max dist | 16 pc | | | 13 pc | | | 11 pc | | |
| $t_{tel}$ | >> 5 yrs | | | 540 days | | | 277 days | | |
| $t_{travel}$ | 1540 days | | | 737 days | | | 550 days | | |
| $t_{tot}$ | >> 5 years | | | 3.5 years | | | 2.3 years | | |

**IWA = 65 mas**

| | dM = 25 mag | | | dM = 26 mag | | | dM = 27 mag | | |
|---|---|---|---|---|---|---|---|---|---|
| | $N_{targets}$ | $\sum_n C_n$ | $t_{det}$ (d) | $N_{targets}$ | $\sum_n C_n$ | $t_{det}$ (d) | $N_{targets}$ | $\sum_n C_n$ | $t_{det}$ (d) |
| A stars | 0 | 0 | 0 | **0** | **0** | **0** | 2 | 0.2 | 2 |
| F stars | 7 | 0.7 | 38 | **5** | **1.7** | **11** | 5 | 3.0 | 11 |
| G stars | 39 | 9.8 | 305 | **20** | **11.3** | **55** | 16 | 11.8 | 34 |
| K stars | 43 | 16.8 | 282 | **24** | **15.2** | **53** | 18 | 13.2 | 24 |
| M stars | 6 | 1.6 | 39 | **3** | **0.9** | **3** | 3 | 0.9 | 3 |
| TOTALS | 95 | 29 | 664 | **52** | **29** | **122** | 44 | 29 | 74 |
| max dist | 14 pc | | | **11 pc** | | | 9 pc | | |
| $t_{tel}$ | 1186 days | | | **320 days** | | | 213 days | | |
| $t_{travel}$ | 1045 days | | | **572 days** | | | 484 days | | |
| $t_{tot}$ | 6.1 years | | | **2.4 years** | | | 1.9 years | | |

**IWA = 50 mas**

| | dM = 25 mag | | | dM = 26 mag | | | dM = 27 mag | | |
|---|---|---|---|---|---|---|---|---|---|
| | $N_{targets}$ | $\sum_n C_n$ | $t_{det}$ (d) | $N_{targets}$ | $\sum_n C_n$ | $t_{det}$ (d) | $N_{targets}$ | $\sum_n C_n$ | $t_{det}$ (d) |
| A stars | 0 | 0 | 0 | 0 | 0 | 0 | 2 | 0.2 | 2 |
| F stars | 2 | 0.3 | 5 | 4 | 1.4 | 8 | 3 | 1.7 | 5 |
| G stars | 20 | 7.5 | 64 | 15 | 9.5 | 33 | 12 | 9.2 | 20 |
| K stars | 32 | 18.4 | 117 | 21 | 15.8 | 35 | 19 | 15.7 | 28 |
| M stars | 7 | 2.8 | 18 | 4 | 2.0 | 3 | 4 | 2.0 | 4 |
| TOTALS | 61 | 29 | 204 | 44 | 29 | 79 | 40 | 29 | 59 |
| max dist | 11 pc | | | 9 pc | | | 9 pc | | |
| $t_{tel}$ | 486 days | | | 228 days | | | 171 days | | |
| $t_{travel}$ | 671 days | | | 484 days | | | 440 days | | |
| $t_{tot}$ | 3.2 years | | | 2.0 years | | | 1.7 years | | |

### 2.7.3 Five Year Mission Performance: Analytical Calculations

The ability of the baseline NWO to reach the required number of targets and cumulative completeness shown in Table 5 within the time allocated for habitable exoplanet observations depends on the total accumulated integration time for detection and characterization plus slew time, as estimated in Equation (20). To gain a back-of-the-envelope estimate of the NWO baseline performance and to illustrate how exozodiacal light impacts mission performance, we used a simple analytical code to produce a target list for each exozodi level by choosing stars in order of decreasing merit value, $F_{merit}$, until the total integration time was 1.5 years or the total mission time was 5 years, whichever came first. We did this for exozodi values of 1-100 zodis, in steps of one zodi. The procedure was performed for three values of $\eta_\oplus$ (0.1, 0.5, and 1.0).

The results for $\eta_\oplus = 0.1$ are displayed in **Figure 15**, which shows how the mission performance is affected by increasing exozodi level. The number of HZs searched (calculated by summing the completeness over all targets) falls off steeply at first (losing ~0.7 HZs per zodi between 1-20 zodis), but becomes nearly constant at higher exozodi levels (losing only ~0.1 HZs per zodi between 60-100 zodis). Although not shown in Figure 15, the average distance of the targets declines in a similar way, showing that as exozodi levels increase the most distant targets are quickly "lost" from the sample target list due to long exposure times. This can be understood by referring back to Figure 11, which shows that the more distant the star, the faster the time increases with increasing exozodi level because the physical area encompassed by the PSF (and the amount of exozodi background flux contained in it) grows with increasing distance. At low exozodi levels, exposure times for all stars are shorter and more distant stars can be observed within the mission lifetime. At high exozodi levels, only the closest targets (i.e., those with shortest exposure times and highest completeness) are observed.

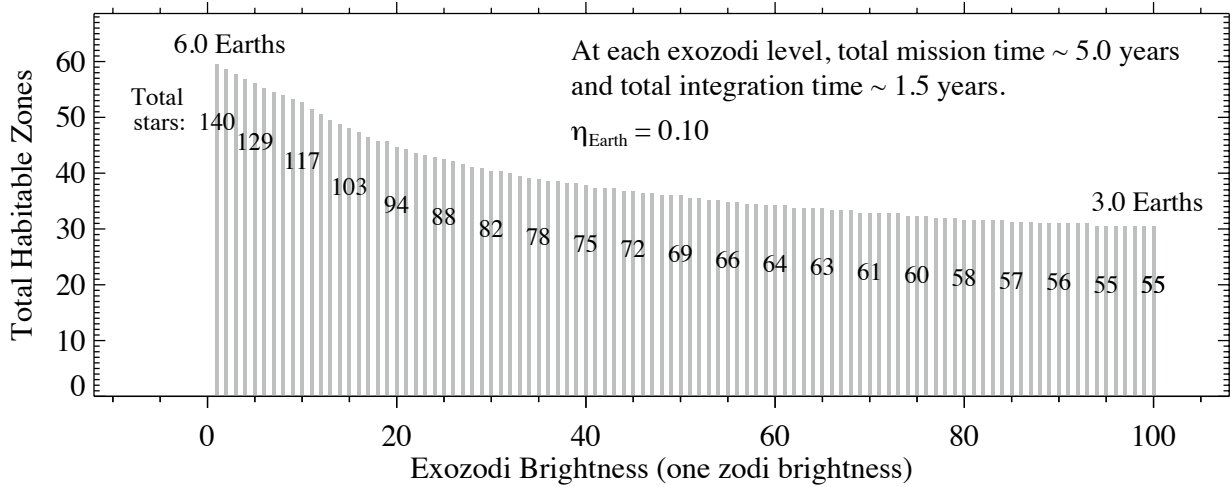

**Figure 15.** Total Habitable Zones searched versus exozodi brightness for the NWO baseline mission, assuming $\eta_\oplus = 0.1$. Stars were chosen in order of decreasing merit until the total on-target integration time reached 1.5 years or the total mission time reached 5 years, assuming 11 days per target for moving the starshade and that we obtained a spectrum with S/N ≥ 10 and R=100 in addition to the imaging observation. Numbers superimposed on the grey bars show the number of target stars observed for various exozodi levels. The total number of Earths characterized is $\eta_\oplus \times$ total HZs searched.

For the case of $\eta_\oplus = 0.1$ shown in Figure 15, the number of stars observed drops to ~55 at the highest exozodi levels, still above the $N_{targets} = 52$ requirement in Table 5. This result illustrates that the NWO baseline specifications are very close to what is required to satisfy the Table 1 objectives. The total completeness achieved in this scenario is ~30 habitable zones, giving an expectation value for the number of planets found of $\langle N_{php} \rangle \sim 3$. Performance results for the other values of $\eta_\oplus$ are given in **Table 7**. The drop in the number of habitable zones searched with increasing exozodi levels is more pronounced at higher values of $\eta_\oplus$, but the number of earths discovered and characterized still improves overall.

The numbers in Table 7 represent the expectation value for the number of planets detected in the scenario where each target is visited once in order of decreasing priority, given the stated assumptions about planet prevalence, planet size and albedo, HZ location, and exozodiacal background. There are inherent uncertainties in these calculations arising from (1) the astrophysical uncertainties such as the distribution in planet parameters and the true extent of the habitable zone as a function of spectral type and planet composition, and (2) the scheduling of targets, which will not always happen in order of $F_{merit}$ due to solar avoidance constraints, $\Delta$V efficiency considerations, and interruptions to the program when interesting objects are found. Nevertheless, for our simplified scenario, Table 7 does illustrate the range in expected results for reasonable values of $\eta_\oplus$ and $\varepsilon$. Given a more detailed understanding of planetary and atmospheric physics, future work could refine our calculations by repeating this process for a range in radius and albedo for planets in the habitable zones of stars of various spectral types.

Note that in Table 7, the number of Earths characterized is the same as the number of Earths detected; i.e., it is assumed that each detection will immediately be followed by spectroscopy. In reality, due to solar avoidance constraints, it is likely that some detections would have to wait for a return visit before spectroscopy ensues. Again, the amount of starshade $\Delta$V limits the total number of targets that can be acquired to about 120 (see §2.7.4), but according to Table 7 this is a limiting factor only for the lowest exozodi brightness values ($\varepsilon = 1$) where the mission objectives are most easily met by the baseline NWO.

Table 7
Calculated Mission Performance Results After 5 Years

| $\eta_\oplus$ | Stars Observed | | | HZs Searched | | | Earths Characterized | | |
|---|---|---|---|---|---|---|---|---|---|
| | $\varepsilon = 1$ | $\varepsilon = 10$ | $\varepsilon = 100$ | $\varepsilon = 1$ | $\varepsilon = 10$ | $\varepsilon = 100$ | $\varepsilon = 1$ | $\varepsilon = 10$ | $\varepsilon = 100$ |
| 0.10 | 140 | 117 | 55 | 60 | 53 | 31 | 6 | 5 | 3 |
| 0.25 | 133 | 90 | 41 | 58 | 43 | 24 | 15 | 11 | 6 |
| 0.50 | 124 | 71 | 32 | 55 | 37 | 19 | 28 | 18 | 10 |
| 1.00 | 107 | 56 | 25 | 50 | 31 | 15 | 50 | 31 | 15 |

The exposure times and habitable zone coverage for the 117 stars selected in the case of $\eta_\oplus = 0.1$ and $\varepsilon = 10$ are displayed in **Figure 16**. For all values of $\eta_\oplus$ and $\varepsilon$, most of these high priority stars are G and K stars, although a handful of A, M, and giant stars are also chosen (note that proxima Cen stands out as having very high completeness for an M star, due to its proximity).

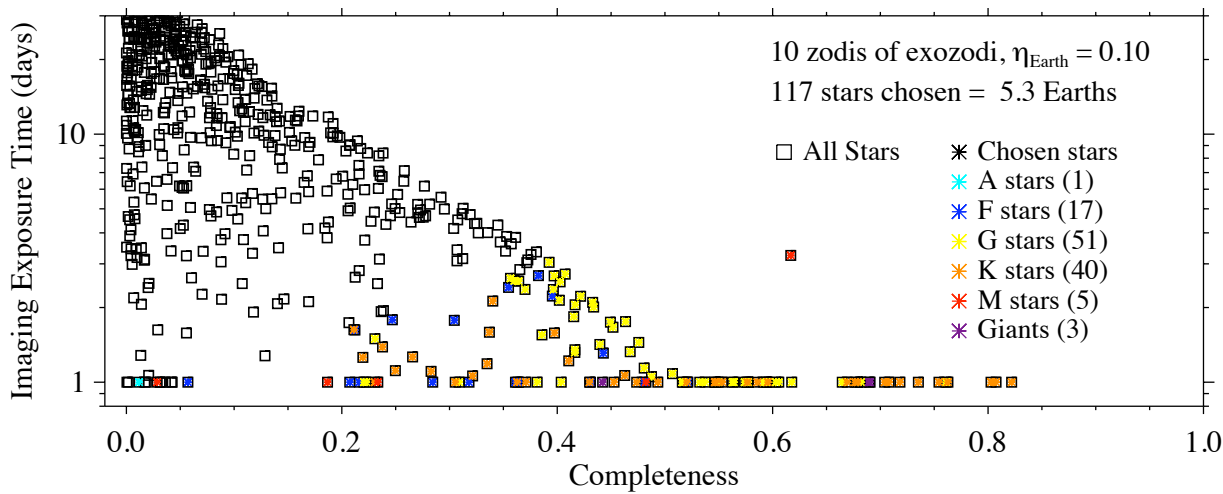

**Figure 16.** Stars chosen according to merit assigned in Equation 21. All stars are plotted with black squares while colored asterisks show the spectral types of chosen targets. For each spectral type, the number chosen appears in parentheses to the right of the type label. The stars were chosen using the method described in the text, assuming $\eta_\oplus = 0.1$, IWA = 65 mas, and $\varepsilon = 10$. In this case, 117 stars are observed, giving 53 total HZs searched and ~5 Earths characterized.

*The Impact of Background Sources.* In order to design a truly optimal observing program we must address the effects of confusion due to exozodiacal clumps and unresolved background objects. How will mission performance be affected if spectroscopy becomes necessary to distinguish planets from exozodiacal clumps or background objects, as a matter of course?

To evaluate this, we repeated the above analysis assuming that spectroscopy is done on every target acquired by the starshade. In this case, the number of stars observed and the total HZs searched become independent of $\eta_\oplus$; they are always equal to the values found in the bottom row of Table 7. However, the number of Earths characterized still depends upon $\eta_\oplus$. Therefore, in all but our worst case scenario ($\eta_\oplus = 0.1$ and $\varepsilon = 100$), the number of targets acquired still meets the NWO science requirements as quantified in Table 5. For the case of $\eta_\oplus = 0.1$ and $\varepsilon = 100$, the expectation value for the number of planets $\langle N_{php} \rangle$ is >1, but performance declines from 55 targets acquired to only 25 targets acquired. According to Table 5, this is not close to the minimum number of targets (52) to achieve 95% probability of detecting at least one planet, and it is unlikely that any path exists where the required number of HZs can be searched.

This finding suggests that, because of the potentially significant impact of faint unresolved background sources, observational strategies to discriminate between planets and confounding sources like background stars and unresolved galaxies -- in a single visit -- should be developed as part of a preparatory science program. Specifically, NWO should be designed to obtain simultaneous images in several bandpasses. This should have little effect on the mission performance, since imaging times are much shorter than the spectroscopy times. It remains for future work to determine the exact bandpasses best suited for this purpose. We emphasize, furthermore, that realistic simulations of direct exo-Earth observations at optical wavelengths in the presence of non-uniform exozodiacal background and background objects have not yet been done.

### 2.7.4 *Mission Performance: Real Time Scheduling*

The above calculations provide an estimate of NWO's expected performance under different astrophysical conditions, assuming a fixed average slew time and unrestricted choice of targets. The above calculations imply that, given the FPB limit = $4 \times 10^{-11}$, IWA = 65 mas, and a wide range of exozodiacal background levels, the 4-m baseline NWO starshade mission is very close to what is needed to satisfy the science objectives for the stars in the solar neighborhood while still reserving more than 50% of the mission lifetime for GA studies. However, this simple calculation does not account for the true distribution of target stars on the sky or realistic engineering constraints which may make the task harder. The 11-day average starshade slew time also does not account for our ability to choose observing sequences that are more efficient with starshade $\Delta V$ and travel time. In this Section, we show the results of a more realistic scheduling model that calculates each possible slew time and emphasizes shorter slews.

Using the set of Hipparcos stars within 30 pc and their completenesses, imaging exposure times, and spectroscopy exposure times, Glassman et al. (2011) have created a Monte-Carlo simulation to find near-optimal observing schedules for starshade missions. This simulation balances the science priority of each observation against mission costs and mission constraints. The costs and constraints considered include the amount of $\Delta V$, and hence fuel, needed to acquire each target ($\Delta V$ limit=8,000 m/s for retargeting; Glassman et al. 2011) and to maintain alignment during each observation ($\Delta V$ limit=300 m/s limit for stationkeeping), the solar-avoidance restrictions (limiting observations to between 45° and 105° from the Sun), and the mission lifetime (5 years). In the cases considered here (as in Glassman et al. 2011), the simulation was run for a 50-m effective-diameter starshade operating 80,000 km from the 4-m telescope. The current simulation does not include revisits, either to continue searching for planets around a star that was already searched or to further characterize a planet that was already detected. Complicating factors for a "multiple visits" simulation include the time-dependence of additional completeness achieved for a previously observed target, the number and timing of revisits to determine orbital information for a previously detected planet, and the appropriate prioritization of revisited targets vs. new targets (Arenberg, Knutson & Schuman 2005). However, a newly discovered planet would be a highly desirable target for follow-up, and the impact of allowing additional visits should be explored in future work.

The simulation first conducts an imaging observation of the initial target star (we begin with alpha Centauri A, but "Target #1" could be any high priority star) and then randomly determines if a planet was discovered, using a probability given by the completeness multiplied by $\eta_\oplus$. If a planet was discovered, the longer spectroscopic characterization observation is conducted for that star. The simulation then moves to the next star, which is randomly chosen from a probability distribution calculated by weighting the benefits and costs of each possible choice. The weighting factors in this probability distribution are $F_{merit}$, the wait time due to solar avoidance, the retargeting angle (starshade slew time), and the station-keeping $\Delta$V costs for each target (which depends on the orbital position of the telescope). Once the mission lifetime or one of the $\Delta$V limits (retargeting or stationkeeping) is reached, the path is terminated.

In this way, we generated an ensemble of 50,000 psuedo-random (or, "smart random") paths, thus avoiding the necessity of generating all $\sim 10^{200}$ possible paths. The generated paths do *not* represent a range of equally likely outcomes. Rather, each path represents a specific option that can be chosen according to the cost (in terms of time and $\Delta$V) and benefits (in terms of science and/or other programmatic requirements that may arise). Though the paths were chosen in a "smart" way, there are still many $\Delta$V- and time-inefficient paths among this ensemble. Once generated, the ensemble of 50,000 paths can be sorted according to the telescope time used, the number of stars searched, and the total number of HZs searched.

This exercise allows us to assess, given the constraints specified above, the variety of paths that are able to meet the science objectives under a range of astrophysical conditions. If we were to use this code as a tool for *scheduling* observations in real time for an operational mission, the procedure would be to first generate an ensemble of paths (with no characterization observations included initially) and then choose the one path that is expected to provide the best performance (as specified by the deterministic parameters such as P($N_{php} > 0$), cumulative completeness, and $\left\langle N_{php} \right\rangle$). Then, when any event interrupts the planned schedule (such as discovering a promising planet), a new ensemble of paths would be generated starting from the telescope's new location relative to the sun, and the best path in terms of science yield would again be chosen.

For the purposes of the current analysis, we are not scheduling a real-time mission but merely simulating the expected results from such a mission under realistic constraints and a variety of astrophysical conditions. Therefore, instead of interrupting the path and re-planning each time we do a characterization observation (as we would in an operational scenario), each path is generated from start to finish with the characterizations included. This simplification was shown by Glassman et al. (2011) to provide a close approximation to the results of the full real-time simulation. We ran the simulation for all nine combinations of $\eta_{\oplus} = (0.1, 0.5, or 1)$ and $\epsilon = (1, 10, or 100)$, to show how the mission is affected by the various astrophysical scenarios.

The simulations in this Section were run using completeness values based on the method described in Arenberg & Schuman (2006), which contains an error in the distribution of orbital parameters. This error led to the simulation using completeness values for the 200 highest priority targets that were, on average, 8% lower than they should have been, with a range of 13% higher to 27% lower that the correct values. For the higher luminosity stars, the completeness was overestimated while for the lower luminosity stars (less than $\sim 2$ $L_{\odot}$), the completeness was underestimated. We were unable to run new schedules using the correct numbers as inputs, but our analytical calculations revealed to us that the effect on the overall results is quite small. In this Section the numbers of HZs searched, for the astrophysical scenarios reported here, reflect the correct values for the stars chosen by the scheduler. We expect that the scheduler results reported here (in terms of number of planets found and characterized) are slightly pessimistic, but that the Figures are accurate in depicting the range of possible programs for a starshade mission.

In **Figure 17**, we show for each ($\eta_\oplus$, $\varepsilon$) scenario how the ensemble of 50,000 paths is distributed in terms of total completeness achieved. All of these paths meet the engineering constraints, and the colored arrows indicate for each scenario the minimum completeness that must be achieved to reach a 95% probability of detecting at least one planet. In most scenarios, the vast majority of paths do meet the science objectives. The one exception to this, as expected, is for the case of very high exozodi levels and low $\eta_\oplus$ (blue curve in the right-most panel). Even so, about 1700 paths (about 3% of the paths) found by the simulation were able to meet the science objectives. Although the options for target scheduling are much fewer, paths do exist that can be chosen in order to meet the science objectives even under these unfavorable conditions.

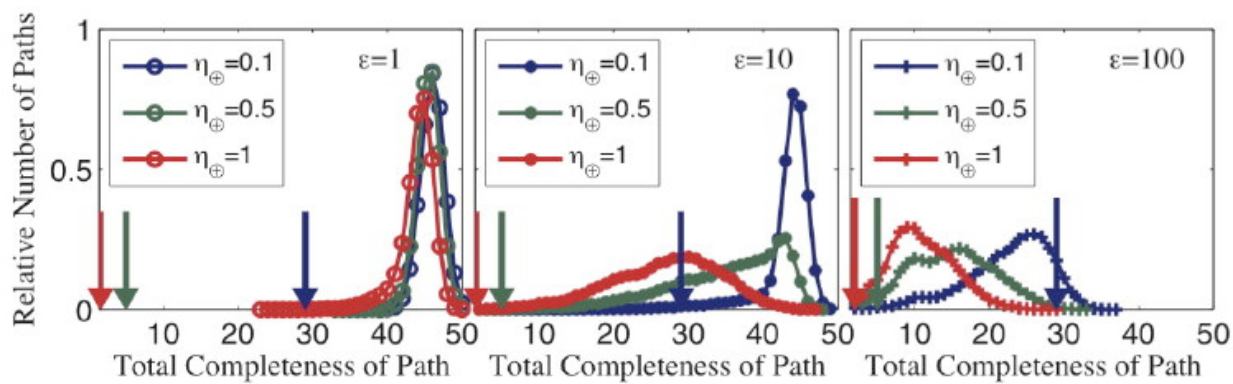

**Figure 17.** The distribution of total completeness achieved (number of HZ searched) for each of nine astrophysical scenarios. The arrows indicate, for each curve, the total completeness that must be attained in order to achieve a 95% probability of detecting at least one planet.

Because the paths in each ensemble include characterization observations, there are certain correlations between the deterministic parameters and randomly drawn parameters that may seem counterintuitive. For example, paths with a higher number of planets discovered ($N_{php}$) have a lower total number of stars searched ($N_{targets}$). This is simply because the characterizations take up time that would otherwise be used for searching more targets. This can be seen in Figure 17 where the total completeness achieved by each path is anti-correlated with $\eta_\oplus$ (and hence with the number of planets characterized), especially at high exozodi levels. However, as shown in Table 5, for higher values of $\eta_\oplus$ it is not necessary to achieve a high total completeness or to search as many targets in order to satisfy the "95% criterion" in Table 1.

*Choosing the "best science" path.* In an operational mission, the path to be executed must be chosen and updated in real time, from among an ensemble of paths generated in a similar manner to that used here. This ensemble of 50,000 paths can be sorted according to deterministic parameters (e.g., cumulative completeness and engineering constraints) to identify paths that meet the science goals even in the more challenging astrophysical scenarios. In the simulated results generated here, the paths can also be sorted by the number of planets detected or characterized, or by the telescope time used, or by the total ΔV consumed for retargeting (8,000 m/s limit) and for station-keeping (300 m/s limit) in order to estimate the science return of the mission.

Glassman et al. (2011) found that the one path yielding the "best" science return (in terms of the number of planets detected and characterized) did not generally correspond to a path with high efficiency in terms of ΔV limits and telescope time used. However, they did find that there are many paths near the 90th percentile science yield that also have very high ΔV efficiencies (~10th percentile in terms of ΔV cost). For the purpose of discussing mission performance, we take the 90th percentile science yield path as our "representative best science path", which is both realistic in terms of science yield and advantageous given the engineering constraints. This path represents the expected results of the mission, assuming that after each interruption to the observing sequence we always calculate and choose the path that (1) has the highest science yield and (2) meets the engineering constraints.

We capture the results of the simulation in **Table 8** for each of the nine scenarios. Table 8 gives the 90th percentile return (from the ensemble of 50,000 paths) of: the number of stars observed, the total number of HZs searched, the number of Earths detected, and the number of Earths characterized for each scenario. The overall performance results, shown graphically in **Figure 18**, indicate that the analytical code used in §2.7.3 does a remarkably good job in characterizing mission performance given a simple ranking of stars and a fixed slew time.

Finally, we note that unlike the simple analytical code used in §2.7.3, the simulation does not assume that spectroscopic characterization is completed immediately after a detection. In these simulations, if after a detection there is not enough time for characterization due to solar avoidance constraints, the simulation simply logs the detection and moves on to the next star. For this reason, the number of characterized Earths is usually less than the number of detected Earths. In real life, the optimal science path will almost certainly include some targets where there is not enough time to complete a full spectroscopic observation, and it may in some instances make sense to carry out at least a partial integration due to the difficulty in re-acquiring planets later on in the mission. The simulations discussed here do not provide for partial spectroscopic integrations or for return visits.

Table 8
Simulated Performance Results: 90th Percentile "Best Science" Path

| $\eta_\oplus$ | Stars Observed | | | HZs Searched | | | Earths Detected | | | Earths Characterized | | |
|---|---|---|---|---|---|---|---|---|---|---|---|---|
| | $\varepsilon = 1$ | $\varepsilon = 10$ | $\varepsilon = 100$ | $\varepsilon = 1$ | $\varepsilon = 10$ | $\varepsilon = 100$ | $\varepsilon = 1$ | $\varepsilon = 10$ | $\varepsilon = 100$ | $\varepsilon = 1$ | $\varepsilon = 10$ | $\varepsilon = 100$ |
| 0.1 | 113 | 107 | 58 | 48 | 46 | 29 | 7 | 7 | 4 | 7 | 5 | 2 |
| 0.5 | 111 | 100 | 43 | 48 | 44 | 23 | 27 | 22 | 11 | 24 | 16 | 7 |
| 1.0 | 108 | 78 | 31 | 47 | 36 | 17 | 47 | 34 | 16 | 43 | 25 | 10 |

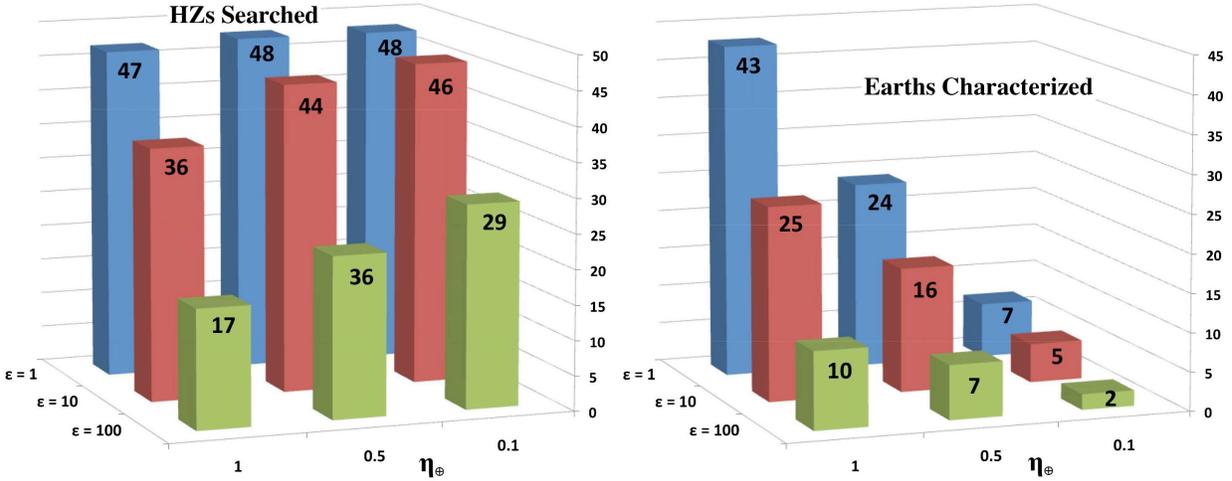

**Figure 18.** Number of habitable zones searched (left) and Earth analogs characterized (right) in simulated observations as a function of $\eta_\oplus$ and $\varepsilon$. The total number of HZs searched is limited by the number of targets that can be acquired during the mission lifetime and completeness per star, thus it is not as strongly affected by $\eta_\oplus$.

**Figure 19** shows the distributions of number of planets detected vs number of stars searched for each of the nine cases. As expected, the number of stars searched is higher for cases with low exozodi levels (since the exposure times are shorter). For the scenarios with the lowest levels of exozodiacal background, starshade $\Delta V$ constraints limit the number of targets that can be acquired to ~120 stars. According to the numbers in Table 5, this indicates that our "limiting $\eta_\oplus$" for a successful baseline mission is $\eta_\oplus \sim 6\%$. At much higher exozodi levels ($\varepsilon=100$), the number of targets is limited to ~60, and the limiting $\eta_\oplus$ is near 8%. Both of these limits could be improved by including revisit searches to high-priority stars.

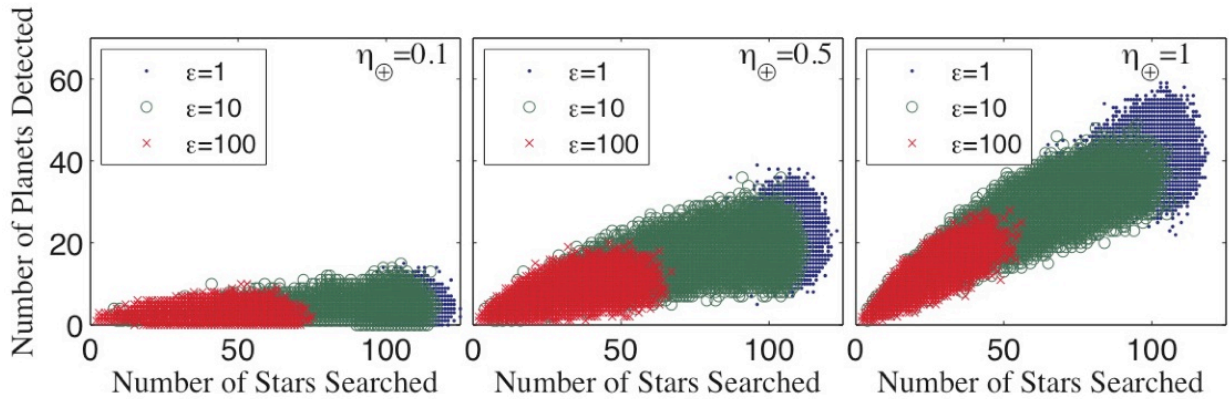

**Figure 19.** The distribution of number of planets detected versus the number of stars searched for each of the nine astrophysical scenarios. The number of stars searched is strongly affected by exozodiacal background levels. For very low levels of exozodiacal light, the number of targets that can be searched is limited to ~120 stars by $\Delta V$ and mission lifetime constraints.

Despite the erroneous completeness values used and other limitations of this code, these results lead to three clear conclusions. 1) For all the astronomical conditions tested, there are many observational schedules that can be found to meet the minimum science requirements, including paths that have better results than the analytical calculations due to shorter starshade travel times. This shows that the starshade design is robust in meeting these requirements, despite the constraints for scheduling the observations. 2) The maximum number of observations that the starshade can conduct is ~120 in a 5 year mission with the current $\Delta$V constraints. This places a limit on the total science return achievable with this mission. 3) Many paths include low priority targets that are located on the sky near to higher priority targets, but the most frequently chosen stars are K and G stars within 15 pc of the Sun. These results are very similar to those from the simple analytical calculations in the previous Section, and most of the stars chosen more than 50% of the time by the scheduler are also shown as "chosen" in Figure 16.

There is much we would like to know about the most favorable NWO targets, and these stars will be examined individually in a future work. The targets represent a diverse cross-section of characteristics, spanning a wide range of ages and metallicities, and several of these systems have known giant planets that fall within the NWO detection space, making them prime targets for comparative planetology. Many of the top targets have stellar companions as well, some of which may have to be ruled out due to stray light contamination, while others may be of questionable merit due to dynamical considerations. Much more detailed analysis will be required to discern whether the "observationally favorable" targets are in fact appropriate for an Earth-finding mission.

## 3. DEFINING THE NWO MISSION

The analysis contained in Section 2 represents the initial steps in determining a viable observing program for the NWO starshade mission. Based on these calculations, it appears that a starshade mission achieving IWA = 65 mas and limiting FPB = $4 \times 10^{-11}$ are appropriate performance requirements for the science goals listed in Table 1, considering (1) high exozodiacal background ($\varepsilon$=1-100 zodis), (2) low planet frequency ($\eta_\oplus$=0.1), (3) realistic $\Delta$V limits, solar avoidance and other logistical constraints, and (4) the true layout of stars in the solar neighborhood, while still reserving 50-70% of the mission time for a rich program of general astrophysics and cosmology. The trade-offs and scaling relations in starshade optical performance have been explored in several recent papers (see Kasdin et al. 2010; Shaklan et al. 2010; Glassman et al. 2010; Glassman et al. 2009); here we describe the physical parameters of a starshade and telescope that would be necessary to achieve the NWO baseline capability.

### 3.1 Starshade and Telescope Parameters
The exact size, shape, and distance of the NWO starshade must be optimized for the required inner working angle, starlight suppression, spectral bandpass, and mission lifetime (Arenberg et al. 2008; Vanderbei et al. 2007). Larger starshades provide additional contrast at longer wavelengths and smaller inner working angles, but must be placed at greater distances. This comes at the expense of greater $\Delta$V costs and fewer target observations.

**Figure 20** shows the shape of a 16-petal flower-like starshade, one of a class of petal-shaped functions for a fully opaque starshade that can suppress diffraction of starlight by many orders of magnitude ($10^{10}$ or more; Cash 2006). The 50-m "effective diameter" assumed by Glassman et al. (2010) and used in this paper corresponds to a "tip to tip" diameter of 62 m, and a "valley to valley" diameter of 29 m. A larger number of starshade petals increases the size of the shadow at the telescope, but this comes at the expense of increased mass and increased scattering of sunlight off the edges of the starshade. Given the 65 milliarcsecond IWA and ~$4 \times 10^{-11}$ limiting FPB proposed in §2, these trade-offs point to a "sweet spot" in the observatory design near a 50-m starshade effective diameter, with 12 or 16 petals, at a distance of 70,000-100,000 km.

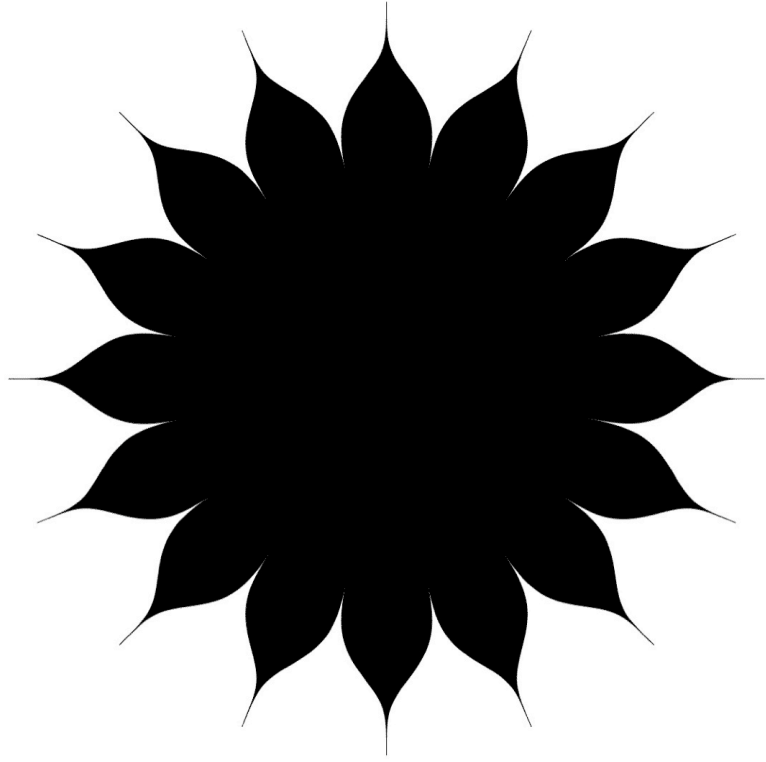

**Figure 20.** A 16-petal starshade as calculated by Cash (2006). The starshade considered in this paper has an effective diameter of 50 m, which corresponds to a "tip to tip" diameter of 62 m, and a "valley to valley" diameter of 29 m.

In **Figure 21** we show the starshade size and distance required to achieve starlight suppression sufficient for a $4 \times 10^{-11}$ limiting FPB (solid lines) and to achieve a given IWA (dashed lines). For the required IWA and wavelength, we select a starshade size and distance near the intersection of the corresponding dashed and solid lines. Any values to the upper right of that point (shaded areas) can achieve both requirements. In the smallest shaded region at right, spectroscopic detection of $O_2$ at 0.76 µm is possible for these planets; in the larger region, only detection is possible, without $O_2$ measurements. With the same 50-m effective diameter starshade at 200,000 km, exoplanets could be detected at 25 mas for the very shortest wavelength (λ = 0.3 μm); or at 120,000 km, exoplanets could be detected at 42 mas for λ ≤ 0.5 μm; or at 40,000 km, exoplanets could be characterized at λ ≤ 1.6 μm for IWA ≈ 130 mas.

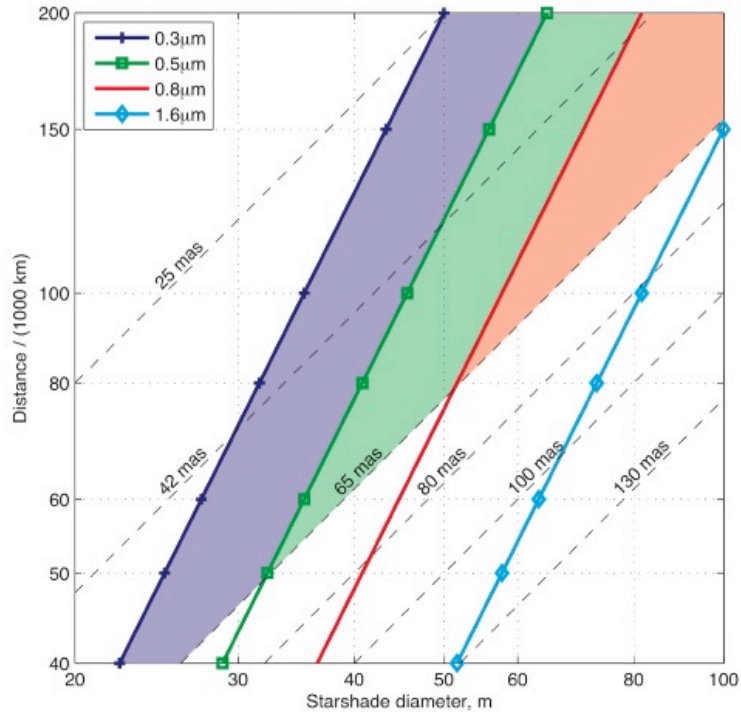

**Figure 21.** The starshade effective diameter and distance required to achieve a $10^{-10}$ starlight suppression at a given wavelength, which we estimate is sufficient to observe planets with FPB = $4 \times 10^{-11}$ (solid lines) and to achieve a given inner working angle (dashed lines). For a given starshade size and distance, starlight suppression improves at shorter wavelengths (as seen in Figure 19b), so the starshade parameters are driven by the longest wavelength in our detection passband. Shaded areas show the range of starshade sizes and distances that will achieve both a 65 mas IWA and $10^{-10}$ starlight suppression at the given wavelength.

### 3.2 *The NWO Baseline Mission Concept*

Using the analysis of §2, we have mapped the objectives in Table 1 onto performance requirements for NWO, and then onto baseline engineering specifications. The NWO architecture outlined in **Table 9** meets or exceeds all of the baseline mission requirements for Terrestrial Planet Finding missions laid out in the 2006 TPF-C Science and Technology Definition report (Levine, Shaklan & Kasting 2006).

The need for a benign gravitational environment suitable for alignment maintenance points to the Earth-Sun second Lagrange point (L2), such as is currently occupied by NASA's WMAP (Bennett et al. 2003), as the optimal orbit location for a starshade mission. This location is also favored in order to minimize stray light sources and maintain high data rates throughout the mission lifetime. The low gravitational and solar-pressure gradient between the two NWO spacecraft at L2 allow the desired mission science to be carried out in a five year lifetime with a reasonable ΔV limits. The launch vehicle, propulsion systems and operations have been described in much greater detail by Cash et al. (2009); a graphical representation of the starshade mission is shown in **Figure 22.**

Table 9
New Worlds Observer Baseline Design

| Science Objectives |
| --- |
| 1. Achieve at least a 95% probability of detecting at least one habitable Earth-like planet, assuming $\eta_\oplus \geq 0.1$ and $\varepsilon \leq 100$, and realistic background sources |
| 2. Characterize spectral features such as water, oxygen and ozone |
| 3. Reserve at least 50% of mission lifetime for General Astrophysics (GA) |

| Constraints |
| --- |
| 1. Complete objectives within a 5 year mission lifetime |
| 2. Maintain a technically feasible mission architecture (with appropriate engineering constraints such as solar avoidance, ΔV limits, etc.) |

| Performance Requirements | |
| --- | --- |
| Wavelength coverage (exoplanets): | 0.3-0.8 μm (detection) |
| Wavelength coverage (exoplanets): | 0.3-1.6 μm (characterization) |
| Photometry (exoplanets): | 4 bandpasses, R~10 |
| Spectral resolution (exoplanets): | R~100 |
| Inner Working Angle (IWA): | 65 mas |
| Planets detectable at 0.8 μm: | $m_p - m_* < 26$ mag |
| Contrast @ planet pixel, 0.8 μm: | $3 \times 10^{-11}$ |
| Angular resolution | 0.026" @ 0.5 μm |
| FOV (exoplanets): | 25" × 25" |
| Wavelength coverage (GA): | 120 nm - 1.7 μm |
| Spectral resolution (GA): | R > 10,000 in the UV |
| FOV (GA): | 10' × 20' (200 sq arcmin) |

| Engineering Specifications | |
| --- | --- |
| Telescope diameter | 4m |
| Starshade separation | 80,000 km |
| Starshade effective diameter | 50 m |
| Number of petals | 16 |
| Mirror coatings | $Al + MgF_2$ or LiF |
| Orbit | Earth-Sun $L_2$ |
| Starshade Lifetime | 5 years |
| Telescope Lifetime | 10 years |
| ΔV budget (targeting) | 8000 m/s |
| ΔV budget (station keeping) | 300 m/s |

### 3.3 *NWO as a General Astrophysics Observatory*

With 50% of the mission lifetime reserved for general astrophysics and cosmology observations, the baseline design for the New Worlds Observer could be considered a flagship astrophysics mission with a "core" science objective to discover and study Earth-like planets orbiting nearby stars. Given UV through near-IR sensitivity, angular resolution nearly a factor of two better than Hubble or JWST, and very high astrometric and photometric stability due to a gravitationally benign and thermally stable environment, the NWO baseline mission would have wide applicability to several fields of study highlighted by the National Academy's 2010 Decadal Survey of Astronomy and Astrophysics, including: (1) exoplanet searches of field stars using transits and lensing, (2) studies of the "cradle to grave" evolution of stars and planetary systems of all masses, (3) coordinated deep synoptic surveys for variable sources such as SNe and GRBs suitable for mapping the distribution of dark matter and the evolution of dark energy, and (4) galactic and extra-galactic astronomy and cosmology to characterize stellar populations, trace the cycling of the ISM, and measure the cosmic evolution of ordinary matter. While it is outside the scope of this paper to describe in detail the instrumental capabilities necessary for such research, the NWO platform seems to represent both the power and the flexibility to unlock a treasure trove of astrophysics from the very nearby to the very distant universe.

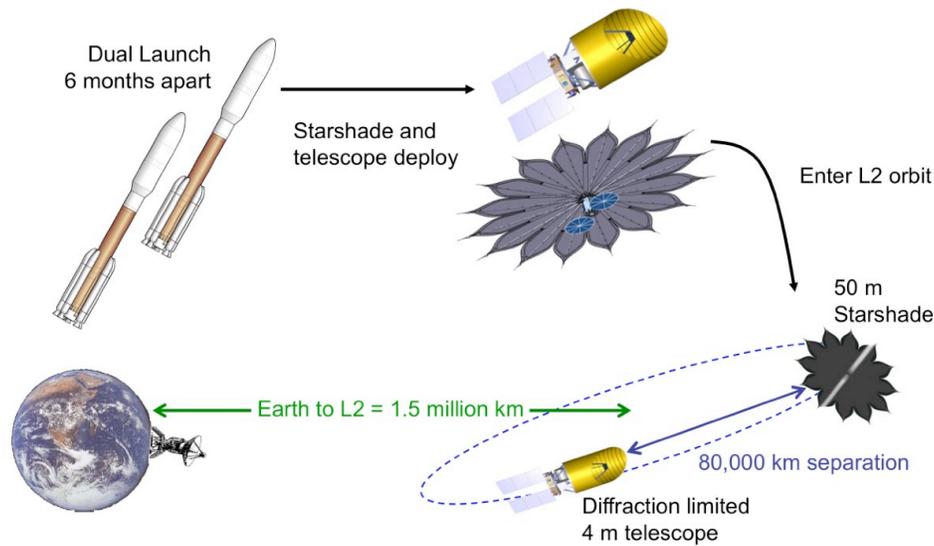

**Figure 22.** Overview of a hypothetical starshade mission. A 4-m telescope and 50-m effective diameter starshade are deployed to Earth-Sun L2 orbit, flying at a separation of 80,000 km for a nominal 5-year mission lifetime.

# 4. CONCLUSIONS AND FUTURE WORK

## 4.1 *Summary*

Our conclusions are as follows:

1. We find that a 4-m NWO with IWA = 65 mas and limiting FPB = $4 \times 10^{-11}$ can achieve a 95% probability of detecting at least one Earth-like planet in a habitable zone within a 5-year mission lifetime, assuming: (1) exozodiacal background levels up to 100 times that of the Solar System, (2) the frequency of planets, $\eta_\oplus$, is 10% or more, (3) reasonable $\Delta V$ and solar avoidance constraints, and (4) the actual distribution of stellar distances, spectral types, and locations on the sky as seen from Earth. This requires just over one year of integration time for detection and characterization, and just under 2.5 years of total mission time (i.e., starshade travel time plus integrations).

2. *De-scoping the mission.* Changing the limiting FPB from 26 to 25 magnitudes significantly degrades the completeness of the highest priority stars, such that many more targets must be observed in order to achieve the "95%" criterion – thereby greatly exceeding the 5 year mission lifetime. On the other hand, relaxing the IWA to 80 mas from the baseline value of 65 mas is not necessarily fatal to the mission. The total integration time required is significantly increased due to loss of HZ visibility for several nearby K and M stars. These targets are replaced by several slightly more distant G stars, but more than 50% of the telescope time is still available for other observations and the total mission time still falls within the 5 year limit.

3. Changing the detection limits affects the completeness of different spectral types such that a deeper FPB will prefer earlier spectral types, while a smaller IWA will prefer later spectral types. However, this effect is somewhat mitigated when target priority is inversely proportional to integration time because the nearest K stars with high completeness are always chosen first. "Custom" target lists that are not dominated by K stars and have a larger number of G stars are possible by factoring spectral type into the priority ranking, but because G stars are on average slightly more distant than K stars, there is limited wiggle room for customizing the target list at the highest exozodi levels within a 5 year mission.

4. With the baseline NWO architecture explored here, the most frequently chosen targets are G and K stars within 15 pc of the sun. Low priority stars located near to high priority targets were sometimes chosen by the scheduler as well. The total number of targets is limited to ~120 stars due to reasonable $\Delta V$ limits for targets acquisition and station keeping.

5. The trade-offs between starshade size, mass, and scattering of sunlight off of the starshade edges all point to a "sweet spot" in the NWO starshade design near a 50-m effective starshade diameter, with 12 or 16 petals, at a distance of 70,000-100,000 km from the telescope. A 4-m telescope appears to be an appropriate baseline for achieving the necessary S/N for planet detection and characterization in a feasible mission lifetime. Such a system is compatible with UV observations and a wide range of General Astrophysics and cosmology programs accounting for 50% or more of the NWO 5 year mission.

## 4.2  *The Way Forward*

### 4.2.1  *Preparatory Science*

The most important factors to be investigated as preparatory science for a New Worlds Observer (or indeed any TPF-like) mission are (1) simulating the detection of planets in the presence of non-uniform and potentially non-grey exozodiacal background, (2) exploring strategies to distinguish between planets and unresolved background galaxies at high galactic latitude, and (3) calculating the contribution of stray light from optical or physical companions, including the effects of changing separations, multiple companions, speckles, and how dithering under the starshade may mitigate some of these concerns.  Finally, observing campaigns to directly measure the faint background that these high proper motion systems will overlay in the next decade may be essential in preparing for a NWO mission.

### 4.2.2  *Engineering "Tall Poles"*

While most of the core starshade technologies are mature, integrating these technologies into a coherent whole needs development and demonstration (Cash et al. 2009).  The "engineering tall poles" for a starshade mission include:

1. Starshade Optical Performance:  The first critical technology is the starshade itself.  Modeling the precise performance of the NWO starshade, correlating these models with testbed results, deriving hardware tolerances, and creating a plan to verify that tolerances are met are critical tasks for the near term.  These results will, in turn, drive requirements on other technologies including shape control, precision deployment, and stray-light control.

2. Starshade Precision Deployment and Shape Maintenance:  The simulation and testbed results discussed in (1) above will define tolerances on thermal distortions, on-orbit dynamics, and mechanical manufacturing error, and other error sources.  Current NWO engineering work involves designing a deployment method that will accommodate the launch vehicle size and mass requirements, deploy the starshade to the required shape, and maintain this shape to the required tolerances despite the various possible mechanical error sources and environmental stressors.

3. Trajectory and Alignment Control:  The second critical technology is the system used to achieve and maintain alignment between the two spacecraft.  This involves two hardware components: (1) an astrometric sensor, which guides the trajectory of the starshade from one star to the next, and (2) a shadow sensor, which provides high-precision alignment measurements while the telescope is in the shadow of the starshade.  For the shadow sensor, a major uncertainty is the long-wavelength diffraction from the starshade that will be used as an input signal, and algorithms to estimate the off-axis position in the shadow using instrument data must be developed and validated.  While most of the technologies for the above tasks are well within the state of the art, there is no existing instrument that can meet all of the requirements.

While many questions remain, these NWO preparatory science and engineering studies will surely advance our understanding of the stars and planetary systems in the solar neighborhood, the possibilities for habitable worlds therein, and the hurdles that must be scaled to find them.

## ACKNOWLEDGMENTS

The NWO Team would like to sincerely thank the anonymous referee for many helpful comments which greatly clarified our discussions and lead us to discover the error in previous completeness calculations. Turnbull is grateful to R. Windhorst (ASU) for helpful discussion regarding the galactic and extragalactic background, to A. Anbar for helpful discussions within the astrobiology community, and to Northrop Grumman Aerospace Systems for support of this work. This work was partly funded by the 2008 NASA Astrophysics Strategic Mission Concept Study and by the ASU "Follow the Elements" NAI Team.

## REFERENCES

Arenberg, J., Glassman, T., Lo, A., Benson, S. 2008, SPIE, 7010, 1S
Arenberg, J., Knutson, H., & Schuman, T. 2005, Technology Review Journal, 13, 2, p. 43, (http://www.is.northropgrumman.com/about/ngtr_journal/assets/TRJ-2005/FW/05FW_Arenberg.pdf)
Arenberg, J. & Schuman, T. 2006, in Direct Imaging of Exoplanets: Science & Techniques. Proc IAU Coll #200, ed. C. Aime & F. Vakili (Cambridge, UK: CUP), 411
Backman, D. E. & Paresce, F. 1993, in Protostars and Planets III, ed. E. Levy & J. Lunine (A93-42937, 17-90; Tucson, AZ: Univ. Arizona Press), p. 1253
Beckwith, S. V. W., Stiavelli, M., Koekemoer, A. M., et al. 2006, AJ, 132, 1729
Bennett, C. L., Halpern, M., Hinshaw, G., et al. 2003, ApJS, 148, 1
Bessell, M. S. 2000, PASP, 112, 961
Bradley, J. P., Keller, L. P., Brownlee, D. E., & Thomas, K. L. 1996, M&PS, 31, 394
Brown, R. A. 2004 ApJ 607, 1003
Brown, R. A, 2005, ApJ, 624, 1010
Bryden, G., Beichman, C. A., Trilling, D. E., et al. 2006, ApJ, 636, 1098
Burrows, C. 2003, in The Design and Construction of Large Optical Telescopes, ed. P. Y. Bely, (New York: Springer-Verlag), Appendix D
Cash, W. 2006, Nature, 442, 51
Cash, W., Kendrick, S., Noecker, C., et al. 2009, Proc. SPIE, 7436, 743606
Catling, D. C., Glein, C. R., Zahnle, K. J., & McKay, C. P. 2005, AsBio, 5, 41
Coe, D., Benítez, N., Sánchez, S. F., Jee, M., Bouwens, R., Ford, H. 2006, AJ, 132, 926
Cox, A. N. 2000, Allen's Astrophysical Quantities (4th ed.; New York: Springer)
Currie, T., Kenyon, S. J., Balog, Z., et al. 2008, ApJ, 672, 558
Dent, W. R. F., Walker, H. J., Holland, W. S., Greaves, J. S., 2000, MNRAS, 314, 702
Dermott, S. F., Jayaraman, S., Xu, Y. L., Gustafson, B. A. S., & Liou, J. C. 1994, Nature, 369, 719
Des Marais, D. J., Harwit, M. O., Jucks, K. W., et al. 2002, Astrobiology, 2, 153
ESA 1997, The Hipparcos and Tycho Catalogues European Space Agency SP-1200 (Noordwijk:ESA)
Festin, L. 1998, A&A 336, 883
Flower, P. J. 1996, ApJ, 469, 355
Frogel et al. 1972, PASP, 84, 581
Gillett, F. C., 1986, IRAS Observations of Cool Excess Around Main Sequence Stars, In: Light on Dark Matter. ASSL, 124, D. Reidel Publishing Co., pp. 61-69

Glassman, T., Lo, A., Arenberg, J., Cash, W., & Noecker, C. 2009, Proc. SPIE 7440, 744013

Glassman, T., Johnson, A., Lo, A., Dailey, D., Shelton, H., & Vogrin, J. 2010, Proc. SPIE, 7731, 7731188

Glassman, T., Newhart, L., Voshell, W., Lo, A. & Barber, G. 2011, IEEE Aerospace Conf., 5747419

Golimowski, D. A., Ardila, D. R., Krist, J. E., et al. 2006, AJ, 131, 3109

Goode, P. R., Qiu, J., Yuchyshyn, V., et al. 2001, GeoRL, 28, 1671

Gould, A., Ford, E. B., & Fischer, D. A. 2003, ApJ, 591, L155

Gould, A., Pepper, F. & DePoy, D. L. 2003, ApJ, 594, 533

Greaves, J. S., Holland, W. S., Wyatt, M. C., et al. 2005, ApJ, 619L, 187

Hernandez, J., Hartmann, L., Calvet, N., et al. 2008, ApJ, 686, 1195

Houk, N. 1978, Catalogue of Two-dimensional Spectral Types for the HD Stars, Vol. 2 (Ann Arbor: Univ. Michigan)

Houk, N. 1982, The Michigan Catalogue of Two-Dimensional Spectral Types for the HD Stars, Vol. 3 (Ann Arbor: Univ. Michigan Press)

Houk, N., & Cowley, A. P. 1975, Michigan Spectral Catalog (Ann Arbor: Univ. Michigan, Dep. Astron.)

Houk, N., & Smith-Moore, A. 1988, Michigan Spectral Survey, Vol. 4 (Ann Arbor: Univ. Michigan Press)

Houk, N., & Swift, C. 1999, Michigan Catalogue of Two-dimensional Spectral Types for the HD Stars, Vol. 5 (Ann Arbor: Univ. Michigan, Dept. Astron.)

Kalas, P., Graham, J. R., & Clampin, M. C. 2005, Nature, 435, 1067

Kalas, P., Graham, J. R., Chiang, et al. 2008, Science, 322, 1345

Kasdin, N. J., Spergel, D. N., Vanderbei, R., et al. 2010, in Proceedings of the Conference in the Spirit of Lyot 2010: Direct Detection of Exoplanets and Circumstellar Disks (Paris, France: University of Paris Diderot), ed. A. Boccaletti, 79.

Kasting, J. F., Whitmire, D. P., & Reynolds, R. T. 1993, Icar, 101, 108

Kelsall, T., Weiland, J. L., Franz, B. S., et al. 1998, ApJ, 508, 44

Krist, J. E., Ardila, D. R., Golimowski, D. A., et al. 2005, AJ, 129, 1008

Kuchner, M. J. & Holman, M. J. 2003, ApJ, 588, 1110

Kuchner, M. J. & Stark, C. C. 2010, AJ, 140, 1007

Lawson, P. R. & Dooley, J. A. 2005, JPL Publication 05-5, http://hdl.handle.net/2014/37428

Lawson, P. R., Lay, O. P., Johnston, K. J. & Beichman, C. A. 2007, JPL Publication 07-1, http://hdl.handle.net/2014/40599

Levine, M., Shaklan, S., & Kasting, J. 2006, JPL Document D-34923, http://planetquest.jpl.nasa.gov/TPF/STDT_Report_Final_Ex2FF86A.pdf

Liebert, J., Dahn, C. C., and Monet, D. G. 1988, ApJ, 332, 891

Lunine, J. et al. 2008, AsBio, 8, 875

Montañés Rodriguez, P., Pallé, E., Goode, P. R., et al. 2004, AdSpR, 34, 293

Moran, S. M., Kuchner, M. J., & Holman, M. J. 2004, ApJ, 612, 1163

Oakley, P. & Cash, W., 2009, ApJ, 700, 1428

Oswalt, T. D., Smith, J. A., Wood, M. A., & Hintzen, P. 1996, Nature, 382, 692

Pavlov, A. A., Kasting, J. F., Brown, L. L., Rages, K. A., & Freedman, R. 2000, JGR, 105, 11981

Pirzkal, N., Sahu, K. C., Burgasser, A., et al. 2005, ApJ, 622, 319

Quijada, M. A., Henry, R. M., Madison, T., Boucarut, R., & Hagopian, J. G. 2010, SPIE, 7739, 77392J

Raymond, O. G., Barkhouser, R. H, Conard, S. J. et al. 2000, Proc SPIE, 4139, 137

Roberge, A., Feldman, P. D., Weinberger, A. J, Deleuil, M., & Bouret, J.-C., 2006, Nature, 441, 724

Ruiz, M. T., & Takamiya, M. Y. 1995, AJ, 109, 2817

Schneider, G., Weinberger, A. J., Becklin, E. E., Debes, J. H., & Smith B. A. 2009, AJ, 137, 53

Shaklan, S., Green, J. J., & Palacios, D. M.  2006, Proc SPIE, 6265, 44

Shaklan, S., Noecker, M. C., Glassman, T., et al. 2010, Proc SPIE, 7731, 75

Su, K. Y. L., Rieke, G. H., Stansberry, J. A., et al.  2006, ApJ, 653, 675

Tinetti, G., Meadows, V. S., Drisp, D., et al. 2006, AsBio, 6, 34

Turnbull, M. C., & Tarter, J. C. 2003a, ApJS, 145, 181

Turnbull, M. S., & Tarter, J. C. 2003b, ApJS, 149, 423

Turnbull, M. C., Traub, W. A., Jucks, K. W., et al.  2006, ApJ, 644, 551

Valenti, J. A., & Fischer, D. A. 2005, ApJS, 159, 141

Vanderbei, R. J., Cady, E., & Kasdin, N. J.  2007, ApJ, 665, 794

Windhorst, R. A., Hathi, N. P., Cohen, S. H., et al.  2008, AdSpR, 41, 1965

Williams, D. M., Kasting, J. F. & Frakes, L. A. 1998, Nature, 396, 453

Woolf, N. J., Smith, P. S., Traub, W. A., & Jucks, K. W.  2002, ApJ, 574, 430

Wright, J. T., Fakhouri, O., Marcy, G. W., et al. 2011, PASP, 123, 412